\documentclass[11pt]{article}
\usepackage[utf8]{inputenc}
\usepackage{amsmath,amssymb,amsthm}
\usepackage[english]{babel}
\usepackage[T1]{fontenc}
\usepackage{graphicx}
\usepackage{url}
\usepackage{multirow}
\usepackage[margin=1in]{geometry}
\usepackage{natbib}
\newcommand{\at}[2][]{#1|_{#2}}
\usepackage{hyperref}
\usepackage[labelsep=period]{caption}
\begin{document}
\title{Dynamic Term Structure Models for SOFR Futures}
\author{Jacob Bjerre Skov and David Skovmand}
\maketitle

\begin{abstract}
The LIBOR rate is currently scheduled for discontinuation, and the replacement advocated by regulators in the US is the Secured Overnight Financing Rate (SOFR). The change has the potential to disrupt the \$200 trillion market of derivatives and debt tied to the LIBOR. The only SOFR linked derivative with any significant liquidity and trading history is the SOFR futures contract, traded at the CME since 2018. We use the historical record of futures prices to construct dynamic arbitrage-free models for the SOFR term structure. The models allow you to construct forward-looking SOFR term rates, imply a SOFR discounting curve and price and risk and risk manage SOFR derivatives, not yet liquidly traded in the market. We find that a standard three-factor Gaussian arbitrage-free Nelson-Siegel model describes term rates very well but a shadow-rate extension is needed to describe the behaviour near the zero-boundary. We also find that the jumps and seasonal effects observed in SOFR, do not need to be specifically accounted for in a model for the futures prices. Finally we study the so-called convexity correction and find that it becomes significant beyond the 2 year maturity. For validation purposes we demonstrate that our model aligns closely with the methodology used by the Federal Reserve to publish indicative SOFR term rates.
\end{abstract}

\section{Introduction}
In 2017 it was announced by the Financial Conduct Authority in the UK that their intention was to phase out LIBOR by the end of 2021. The same year the Alternative Reference Rates Committee (ARRC) announced that the recommended replacement in the US, would be the Secured Overnight Financing Rate (SOFR). The SOFR is an overnight rate based on repo transactions for US treasury securities. The rate was chosen due to the sizeable volume of the repo market, but also because it was not a direct policy rate, unlike the EFFR, which is the rate the Fed refers to in the published targets of the Federal Open Market Comittee (FOMC). SOFR has already been partially implemented, replacing the EFFR as the the main rate used for discounting\footnote{The switch from EFFR to SOFR as the primary discounting rate at the LCH and CME clearing houses took place on October 16 2020.}. This is part of a global trend as regulators throughout the world have pushed towards moving away from LIBOR or its equivalents to overnight transactions based rates generically referred to as a Risk-Free Rates (RFR)\footnote{In the UK SONIA, the Eurozone has chosen ESTER, SARON in Switzerland, and so on.}.
\\ \indent
The transition from LIBOR to SOFR is not frictionless as SOFR is fundamentally different from LIBOR. The key difference is that LIBOR is reported across multiple tenors covering term lending up to 12 months. SOFR, being an overnight rate, by construction has no in-built view of the future beyond the 24 hour term. Furthermore, SOFR has shown itself to be extremely volatile at times. This was exemplified by the September 17, 2019 so called "SOFR Surge" where the rate jumped 282 basis points compared to the previous day. The ARRC has recommended (see for example \cite{ARRCSAFR}) that LIBOR is replaced by a running three month compounded daily average of SOFR, and this will naturally decrease the volatility. The recommendation for existing contracts is that the LIBOR fixing should be replaced by this backward-looking average rate plus a spread that reflects the historical median spread to LIBOR.
\\ \indent
Changing the fixing rate from LIBOR to some transformation of SOFR in an existing contract can in principle be easily done if both parties of the loan or derivative agree to make the switch\footnote{In practicality there are multiple complications and potential for legal battles see for example \cite{henrard2019libor} for a quant perspective on this part of the transition.}. But moving from EFFR discounting to SOFR discounting requires, not only the value of the benchmark itself, but a liquid market of forward-looking instruments underlying SOFR. The previous practice of discounting using EFFR relied heavily on a liquid OIS market to derive the EFFR discounting curve. A similar market for SOFR OIS' does not yet exist. The ISDA counted just 381 trades in all SOFR related swaps (including basis swaps) the week ending February 19 2021. This compared to 14,321 trades in swaps referencing USD LIBOR.
\\ \indent
Having a liquid derivatives market underlying SOFR is desirable not just for deriving a discounting curve, but also for implying the "term SOFR" i.e. a discrete forward rate referencing the markets expectation of SOFR over the term period. Market participants \cite{Risk2020} have called for such a benchmark, preferring the forward-looking nature of LIBOR over the backward-looking SOFR average. As a result, it is written in the transition plan of the ARRC (See \cite{ARRCTransitionplan}) that a term SOFR benchmark rate should be published by the end of 2021. Various methods for constructing such a benchmark have been suggested by the ARRC in \cite{ARRCSecondreport}. They express some doubts that the SOFR swap market will be robust enough in the near future to underlie an index due to its lack of liquidity. This leaves SOFR futures as the only realistic underlying asset class for term SOFR benchmark. The average daily trading volume of SOFR futures has been increasing steadily since their introduction at the CME in 2018 and for the month of January 2021 it was 1,8 mio. contracts up 117\% from January 2020. Translating futures prices into a term structure is inherently a model dependent task due to the so called convexity correction measured as the difference between the forward and futures rate (see for example \cite{hunt2004financial}). Model-free attempts to calculate the convexity generally requires volatility inputs derived from non-linear derivatives, but such a market with a SOFR underlying is not likely to materialize until LIBOR is fully discontinued\footnote{Since Jan 2020 the CME also offers options on SOFR futures. But by January 2021, there has been little to no trading in these instruments.}. This effectively means that using a model is the only way to consistently construct the SOFR term structure. And of course also the only way to price and hedge SOFR derivatives. If the predictions of the regulators come to fruition and SOFR becomes the main benchmark replacing LIBOR, it would be natural to expect a liquid market to spawn of SOFR equivalents to the massively popular LIBOR derivatives such as swaptions, caps, floors and other non-linear derivatives. A model for the dynamics of the SOFR term structure based on futures data would therefore be particularly important to price and hedge these products in the early stages of the transition, when futures are arguably the only credible source for market expectations of SOFR.
\\ \indent
In this paper we study a variety of term structure models for SOFR. Our primary focus is on models that can be easily estimated to futures data, using standard statistical methods. With this view, we start the analysis with Gaussian affine term structure models in particular the arbitrage-free Nelson-Siegel framework of \cite{christensen2011affine}, and its shadow-rate extension in \cite{christensen2016modeling}.
The models are estimated using the maximum-likelihood Kalman filtering method. This is straightforward in the purely Gaussian models, but for the Shadow-rate model it is complicated by a lack of closed form expressions for the futures prices. We develop an approximation formula and verify its accuracy using simulation methods. 
\\ \indent
Our work is to the authors knowledge the first to estimate dynamic arbitrage-free models for the SOFR term structure. Other dynamic models for SOFR have been put forward for example \cite{lyashenko2019looking}, \cite{macrina2020rational}, \cite{andersen2020spike}. But the existing approaches have mainly focused on pricing derivatives on the compounded SOFR average, and have not taken their models to actual data. An exception is \cite{gellert2021short}, that perform a multi-date calibration under the risk-neutral measure with a view towards analyzing the impact of SOFR spikes. 
\\ \indent
The work draws heavy inspiration from \cite{heitfield2019inferring}, who have put forward a model free approach to translating SOFR futures prices into a forward looking term rate. In particular their approach is currently used by the Federal Reserve to publish an indicative term SOFR rate derived from futures prices. We use their indicative rates as a robustness check on our model and find that our model aligns with those rates to a very high degree. We furthermore find that the jumps and seasonal effects observed in SOFR do not need to be specifically accounted for when the goal is to describe the futures curve. 
The authors in \cite{heitfield2019inferring} do not use a dynamic arbitrage-free term structure model, and thus in effect ignore the convexity correction. Convexity has been studied for SOFR futures in \cite{henrard2018overnight} and in more detail in \cite{rosen2019averaged}, but not to the authors knowledge in a model estimated to the historical record. Having an estimated dynamic model allows us to gauge the practical relevance of convexity. We find evidence that the convexity correction, shows significant model dependence at the lower bound, but is only relevant if futures prices are used to build a term structure beyond 2-3 years in the future. In particular the convexity correction is not significant when translating futures prices into term rates with a tenor less than one year which aligns with the methods of \cite{heitfield2019inferring}. Finally we study how the spike in SOFR of September 2019 as well as the March 2020 drop in overall interest rates has impacted the SOFR and the related rates. We find that the Gaussian models and the shadow-rate models yield similar results up until March 2020 rate drop. But we find that Gaussian models are unable to capture the volatility compression that occurs after this rate drop.  
\\ \indent
The paper is structured in the following way. We begin with a general description of our empirical setup beginning with a general description of for forward looking term rates. We describe the models the pricing formulas, the data and estimation method relegating most of the mathematical detail to the appendix. In section 3 we present our results and in section 4 we investigate the size and dynamics of the the convexity adjustment for SOFR futures. 

\section{Empirical Setup}
\subsection{Defining SOFR term rates}
\begin{figure}[hbt!]
\centering
\includegraphics[angle=0, scale=.7]{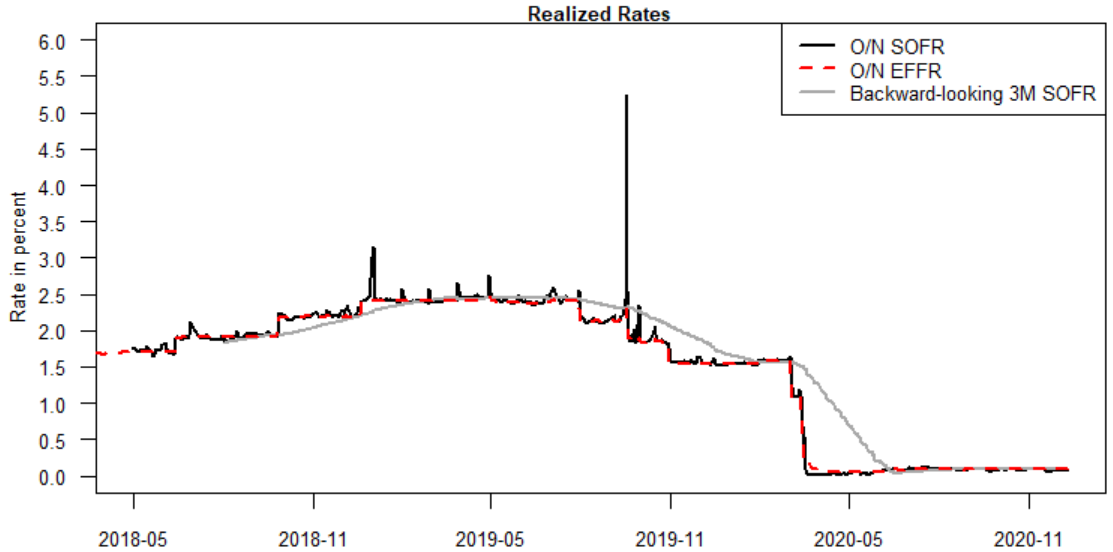}
\caption{Overnight SOFR and EFFR as well as the backward-looking discretely compounded three-month SOFR term rate.}
\label{fig:sofr_back}
\end{figure}

Define the filtered probability space $(\Omega  \mathcal{F},\{\mathcal{F}_t\}_{t \ge 0 },Q)$, with $Q$ being the risk neutral measure defined the by the continuous savings account numeraire given by $B_t=e^{\int_0^t r_s ds}$. $\{r_t\}_{t\geq 0}$ is the risk free short rate. We can then define the zero coupon bond as
\begin{align}
p(t,T)=B_t\mathbb{E}^Q\left[B_T^{-1} |\mathcal{F}_t \right].\label{SOFR}
\end{align}
This allows to construct the discrete overnight SOFR is defined as,
\begin{align}
R_{d_i}(t_i)=\frac{1}{d_{i}}\left(\frac{1}{p(t_i,t_{i}+d_{i})}-1 \right)    
\end{align}
with $d_i$ denoting the day count fraction multiplied by the amount of days to which the overnight rate applies. E.g. $d_i=3/360$ on Fridays and $d_i=1/360$ on business days that are not followed by a holiday. \\
A critical part of the LIBOR transition has been to define a replacement benchmark for LIBOR based on the new RFRs. The regulatory agencies and the ARRC has decided LIBOR is to be replaced by a backward-looking compounded average calculated as 
\begin{align}
R^{B}(S,T)=\frac{1}{T-S}\left(\prod_{i=1}^{N} \left(1+d_i R_{d_i}(t_i)\right) -1 \right).\label{backwardlookingrate}
\end{align}
Unlike LIBOR rates which are fixed in advance the RFR based backward-looking rates are only known at the end of the term and thus $\mathcal{F}_T$-measurable, hence they can hardly be considered as proper term rates. Figure \ref{fig:sofr_back} plots the three month compounded term SOFR term rate against the overnight SOFR and EFFR. The SOFR is closely related to the federal funds rate, however it displays clear rate spikes on certain month, quarter or year ends known from the repo market it is based on. The figure shows that the backward-looking running average is significantly volatility reducing to the point that even the September 2019 spike in SOFR reduces to a minor uptick in the backward-looking benchmark.
\\ \indent
While it cannot be said that the backward-looking rate has been excessively volatile, it has nevertheless been highlighted by the \cite{ARRCSecondreport} that a forward-looking term rate is preferable by market participants, since it would incorporate market expectations about the future rates, but also because of the so called 'measurability problem' (see \cite{henrard2019libor}) that arises from having a rate that is measurable at the end as opposed to the beginning of the term.  However, assuming that a synthetic zero-coupon bond term structure  can be implied from market prices such a proper forward looking term rate is simply calculated as the discrete spot rate,
\begin{align}
R^{F}(S,T)=\frac{1}{T-S}\left(\frac{1}{p(S,T)}-1\right).\label{SOFRtermrate}
\end{align}
The backward and the forward-looking rates are easily seen, by repeated application of the tower property, to be related through discounted expectation as
\begin{align}
B_t\mathbb{E}^Q\left[B_T^{-1} R^{F}(S,T)|\mathcal{F}_t \right]=B_t\mathbb{E}^Q\left[B_T^{-1} R^{B}(S,T)|\mathcal{F}_t \right],\, \quad t\leq S\leq T. \label{forwardbackwardrelation}
\end{align}
Which means that the price of any forward starting linear derivative will be the same at any point before the accrual date regardless of whether a backward or forward-looking benchmark is used to determine the cash-flows. In particular setting $t=S$, and changing to the $T$-forward measure we get
\begin{align}
\label{prediction}
R^{F}(S,T)=\frac{B_S}{p(S,T)}\mathbb{E}^Q\left[B_T^{-1} R^{B}(S,T)|\mathcal{F}_S \right]=\mathbb{E}^T[R^B(S,T)|\mathcal{F}_S],
\end{align}
from which we can see that the forward looking term rate is a prediction of the backward-looking rate under the $T$-forward measure.
\\ \indent
An Overnight-Index-Swap (OIS) is a fixed for floating swap with floating payments equal that of the backward looking rates in \eqref{backwardlookingrate}. From \eqref{forwardbackwardrelation} and \eqref{prediction} it therefore follows that OIS prices would determine forward looking term rate in a model free manner. But as stated above the OIS market with a SOFR underlying is not yet sufficiently liquid. The current market for OIS is still very much dominated by EFFR as the underlying overnight rate. We are thus left with the SOFR futures as the main source of data for inferring forward looking SOFR term rates. Since this is an inherently model dependent task we will consider the ability of a model to construct correct forward looking SOFR term rates from futures data as one of the main criteria to assess the performance of the models specified in the next section. 
\subsection{Constructing the Short Rate models}
When modelling the term structure of futures rates we consider one-, two-, and three-factor versions of Gaussian arbitrage-free short-rate models. Since unrestricted affine multi-factor models often suffer from over-parametrization resulting in multiple maxima of the likelihood function we consider the parameter restrictions of the arbitrage-free Nelson-Siegel (AFNS) model presented in \cite{christensen2011affine} belonging to the class of Gaussian affine term structure models. The AFNS model is a three-factor model effectively replicating the yield factor loadings of the popular Dynamic Nelson-Siegel model in \cite{diebold2006forecasting}. Furthermore, we consider a two-factor version of the AFNS model omitting the final (curvature) factor as well as the standard Gaussian single-factor model by \cite{vasicek1977equilibrium}. Finally, we consider a shadow rate extension of the AFNS model where the short rate satisfies the zero lower bound as in \cite{black1995interest}.
We briefly outline the general setup of the models used in this paper, while model specific parameter restrictions and results are given in appendix A. As before we define the filtered probability space $(\Omega , \mathcal{F},\{\mathcal{F}_t\}_{t \ge 0 },Q)$ and assume that the state variable, $\{X_t\}_{t \ge 0 }$, is a Markov process which solves the stochastic differential equation
\begin{align}
\label{duffie}
dX_t=K^Q[\theta^Q - X_t] dt + \Sigma dW^Q_t,
\end{align}
where $K^Q$ and $\Sigma$ are $N\times N$ matrices, $\theta^Q\in \mathbb{R}^N$ and $\{W^Q\}_{t\geq 0}$ is an $\mathcal{F}$-adapted Brownian motion on $\mathbb{R}^N$. In the Gaussian models the short rate is given by
\begin{align}
r_t=\rho_0 +\rho_1' X_t,
\end{align}
with $\rho_0$ a scalar and $\rho_1 \in \mathbb{R}^N$. While in the shadow rate model we let $s_t=\rho_0 +\rho_1' X_t$ denote the shadow short rate and define 
\begin{align}
r_t=\max(s_t,0)
\end{align}
as the actual short rate. As in the previous section we set  $B_t=e^{\int_0^tr_sds}$ and $p(t,T)=B_t\mathbb{E}^Q\left[B_T^{-1} |\mathcal{F}_t \right].$\\ The real world dynamics are connected to the risk neutral dynamics by specifying a market price of risk, $\Lambda_t$,
\begin{align}
dW^Q_t=\Lambda_t dt + dW^P_t.
\end{align}
We consider an essentially affine market price of risk as in \cite{duffee2002term}, which under Gaussian state variable dynamics reduces to
\begin{align}
\Lambda_t=\lambda_1 +\lambda_2 X_t.
\end{align}
The resulting $P$-dynamics share the same form as the $Q$-dynamics and are thus also Gaussian
\begin{align}
dX_t=K^P[\theta^P - X_t] dt + \Sigma dW^P_t.
\end{align}
Throughout we will assume independent dynamics of the state variables under the physical measure and thus restrict $K^P$ and $\Sigma$ to be diagonal matrices.

\subsection{Pricing SOFR Futures}
%The discrete overnight SOFR is defined as,
%$$R_{d_i}(t_i)=\frac{1}{d_{i}}\left(\frac{1}{p(t_i,t_{i}+d_{i})}-1 \right)$$ with $d_i$ denoting the day count fraction multiplied by the amount of days to which the overnight rate applies. E.g. $d_i=3/360$ on Fridays and $d_i=1/360$ on business days that are not followed by a holiday. 
The settlement price of the futures contract is quoted as $100(1-R(S,T))$ where $R(S,T)$ denotes the futures rate and is a function of the discrete overnight reference rates during the contract period and thus $\mathcal{F}_T$-measurable. $R(S,T)$ is defined differently for One-Month and the Three-Month SOFR futures. In both cases we can use the well known result (see for example \cite{hunt2004financial}) that the value of a futures contract with a random payoff equals the risk neutral expectation of the non-discounted payoff. The time $t$ futures rate between time $S$ and $T$ can therefore be computed as
\begin{equation}
f(t;S,T)=\mathbb{E}^Q\left[R(S,T)|\mathcal{F}_t \right].
\end{equation}
We present general futures rates formulas for both one- and three-month futures contracts. Specific solutions to the AFNS model are provided in appendix A, and the development of the approximation formulas for shadow rate model can be found in appendix B. 
\subsubsection{One-Month SOFR Futures}
The one-month futures rate is based on the arithmetic average of the daily reference rate during the contract month
\begin{align}
R^{1m}(S,T)=\frac{1}{N}\sum_{i=1}^N R_{d_i}(t_i).
\end{align}
Where $N$ denotes the total number of days in the month and $R_{d_i}(t_i)$ for $i\in 1,...,N$ with $S\le t_1,...,t_N\le T$ the published SOFR rates given in \eqref{SOFR}. For any date for which the rate is not published the last preceding rate is used, as specified in the futures contract. As in \cite{mercurio2018simple} we approximate the discrete average by an integral of the instantaneous short rate
\begin{equation}
R^{1m}(S,T)\approx \frac{1}{T-S}\int_S^T r_s ds.
\end{equation}
As we show in appendix D this approximation is strictly not necessary as an exact solution does exist in the affine case. We compute the exact one-month rate in appendix D, which shows that the approximation error is similar across all contracts and only a fraction of a basis point. We nevertheless apply the approximation for ease of computation.\\
Denote the time $t$ rate of the one-month futures starting to accrue at time $S$ and with settlement on time $T$ by $f^{1m}(t;S,T)$ then
\begin{align}
\label{1mapp}
f^{1m}(t;S,T)&=\frac{1}{T-S}\mathbb{E}^Q\left[\int_S^T r_s ds|\mathcal{F}_t\right].
\end{align}
When pricing a futures contract settling at the end of the current month and thus $S<t$, we have to account for the part of the underlying rate that has already been accrued. In this case the futures rate is calculated as
\begin{align}
f^{1m}(t;S,T)
=\frac{1}{N} \sum_{i=1}^{N_0} R_{d_i}(t_i)+\frac{1}{T-S}\mathbb{E}^Q\left[\int_t^T r_s ds|\mathcal{F}_t\right]
\end{align}
where $R_{d_i}(t_i)$ for $i\in 1,...,N_0$ with $S\le t_1,...,t_{N_0}\le t$ are realized rates and thus $\mathcal{F}_t$-measurable.
The settlement of the one-month federal funds futures is based on the same specifications and the pricing formula is therefore valid for both SOFR and federal funds one-month futures traded at the CME.
\subsubsection{Three-Month SOFR Futures}
The three month futures contract is based on the daily compounded reference rate during the contract quarter
\begin{align}
R^{3m}(S,T)=\frac{1}{T-S}\left(\prod_{i=1}^{N} \left(1+d_i R_{d_i}(t_i) \right)-1\right).
\end{align}
Where $R_{d_i}(t_i)$ for $i\in 1,...,N$ with $S\le t_1,...,t_N\le T$ denotes the realized overnight rates in the reference quarter and $d_{t_i}$ the amount of days to which $R_{d_i}(t_i)$ applies.
Again, we follow \cite{mercurio2018simple} and approximate the daily compounded rate by the continuously compounded rate
\begin{equation}
R^{3m}(S,T)\approx \frac{1}{T-S}\left(e^{\int_S^T r_s ds}-1\right).
\end{equation}
The accuracy of the continuous approximation is studied in appendix D showing errors of an order less than $10^{-7}$ for all open contracts. This again is applied mainly for ease of computation.
The time $t$ rate of the three-month future starting to accrue at time $S$ and with settlement on time $T$ is then given by
\begin{equation}
\label{3mapp}
f^{3m}(t;S,T)=\frac{1}{T-S}\left(\mathbb{E}^Q\left[e^{\int_S^T r_s ds}|\mathcal{F}_t \right]-1\right). 
\end{equation}
When $S<t$ and part of the underlying rate has already accrued we account for this in the pricing formula using discrete compounding
\begin{align}
f^{3m}(t;S,T)=
\frac{1}{T-S}\left(\left(\prod_{i=1}^{N_0} \left[1+d_i R_{d_i}(t_i) \right]\right)\mathbb{E}^Q\left[e^{\int_t^T r_s ds}|\mathcal{F}_t \right] - 1 \right)
\end{align}
where $R_{d_i}(t_i)$ for $i\in 1,...,N_0$ with $S\le t_1,...,t_{N_0}\le t$ denotes $\mathcal{F}_t$-measurable realized overnight rates.
\subsection{Data and Estimation}
To infer SOFR-based term rates we estimate the models using end of day prices on CME SOFR futures contracts. All futures data is collected through Refinitiv Eikon. The models are fitted using the extended Kalman filter maximum likelihood method described in appendix B. Code for the extended Kalman filter and relevant pricing functions is available and uploaded at Github\footnote{\url{https://github.com/Jacob-Skov/DTSM-SOFR-Futures}}. We apply historical futures data starting from the 19th of June 2018 up until the day where the term rates are computed. Term rates are based on a minimum of 250 historical daily futures data observations allowing us to compute term rates from June 2019 and onwards. We calculate the term rates using equation \eqref{SOFRtermrate}, applying the ACT/360 day count convention and the modified following business day convention on the USNY business day calendar.
\\ \indent
The quality of the obtained model estimates and resulting term rates strongly rely on the observed futures prices accurately reflecting market expectations on futures rates. As shown in \cite{heitfield2019inferring} the majority of the liquidity in the SOFR futures markets is concentrated around the nearest one-month contracts and three-month contracts at a one year horizon. Based on this we use observations of one-month contracts for the seven nearest calendar months as well as the five closest quarterly contracts. The data therefore reflects market expectations covering a period of just over a year. Daily term rates are then calculated using the latest set of filtered state variables together with the optimal set of parameters provided by the Kalman filter. In appendix B we perform a simulation study testing the efficiency of the Kalman filter maximum likelihood method. The study shows that the parameters related to the risk neutral dynamics and thus determining the term structure are accurately identified by the filter showing no significant bias. It should further be noted that while we construct term rates using end of day quotes in this paper, one can apply the same approach at any point during the day using the Kalman filtering on the prevailing futures prices while using e.g. end of day historical data in the estimation to obtain a relevant term structure.
\section{Empirical Results}
\begin{figure}[hbt!]
\centering
\includegraphics[angle=0, scale=.7]{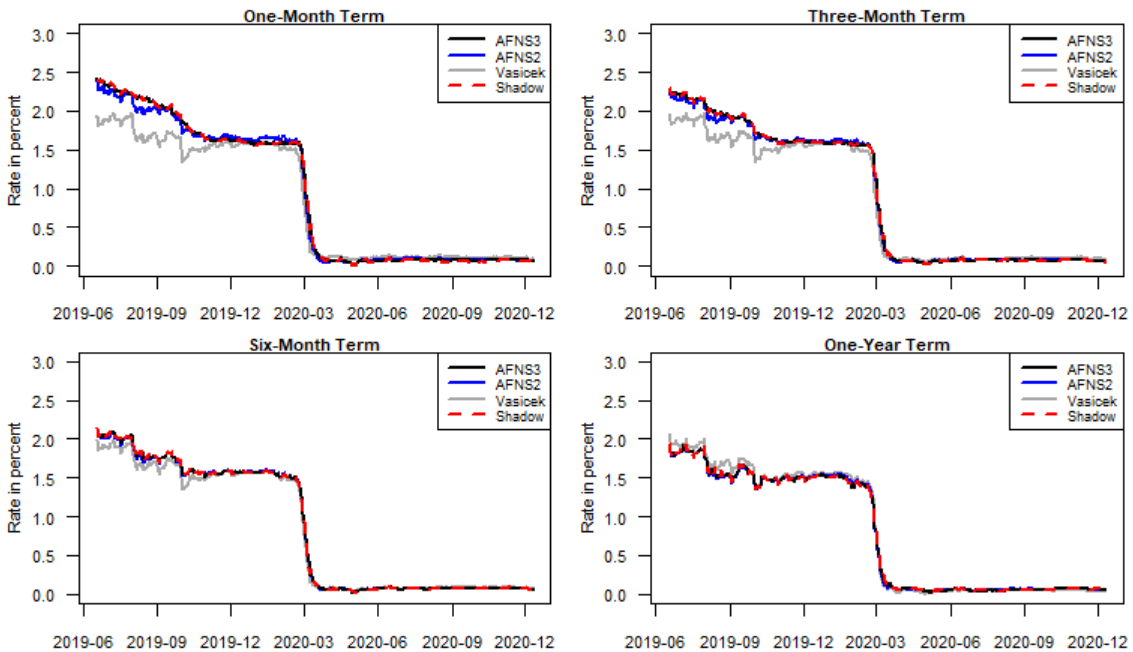}
\caption{Model implied arbitrage-free term rates for different tenors based on rolling re-estimates from 17th of June 2019.}
\label{fig:sofr}
\end{figure}

\begin{figure}[hbt!]
\centering
\includegraphics[angle=0, scale=0.63]{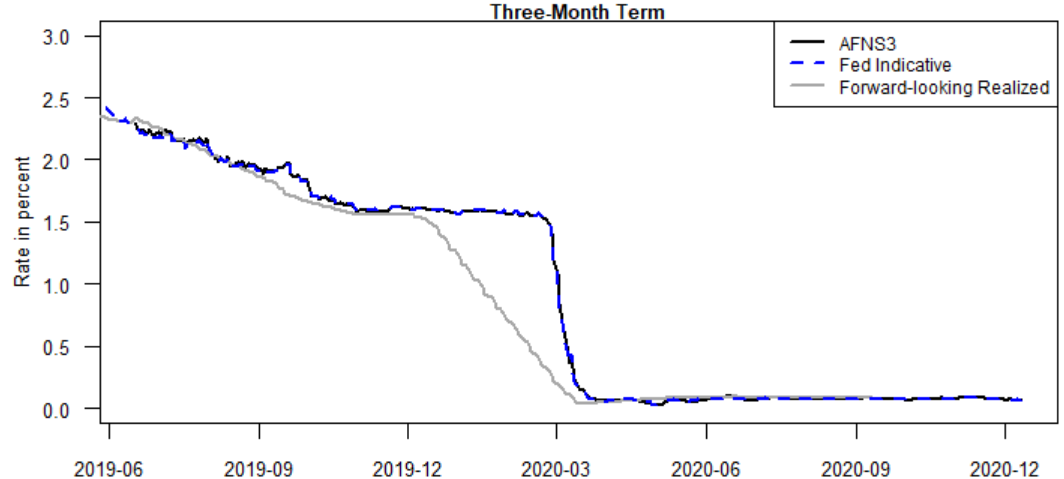}
\caption{Comparison of forward-looking three-month SOFR term rates.}
\label{fig:sofr_comp}
\end{figure}

\begin{figure}[hbt!]
\centering
\includegraphics[angle=0, scale=0.63]{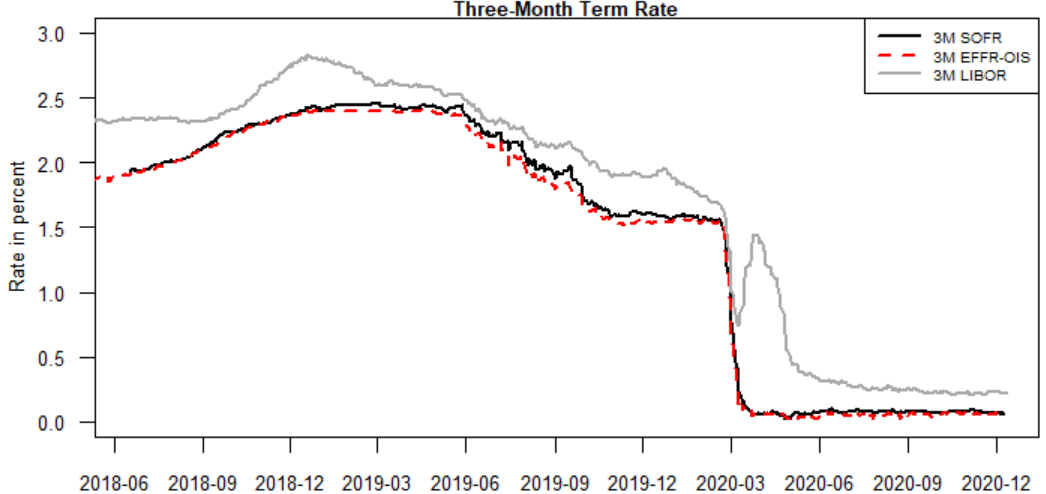}
\caption{Comparison of forward-looking three-month term rates. The three-month SOFR term rated is based on the final AFNS estimates using the set of filtered state variables.}
\label{fig:libor_comp}
\end{figure}

Figure \ref{fig:sofr} presents the SOFR-based term rates obtained using the one-, two-, and three-factor Gaussian models. Particularly at the one-month tenor the models show significant variation in the model implied term rates. Also the graphs show that when market expectations of futures rates are stable implying a flat yield curve all three models provide near identical term rates. This can be seen during the most recent period after the drastic lowering of rates following COVID-19 due to announcements made by Federal Reserve to keep rates near zero for an extended amount of time.
\\ \indent
\begin{table}[hbt!]
\small
\center
\begin{tabular}{ccccccccccccc}
\textbf{Model}  & \multicolumn{7}{c}{\textbf{One-Month}}  & \multicolumn{5}{c}{\textbf{Three-Month}}                      \\ \hline 
                                & \textbf{1st} & \textbf{2nd} & \textbf{3rd} & \textbf{4th} & \textbf{5th} & \textbf{6th} & \textbf{7th} & \textbf{1st} & \textbf{2nd} & \textbf{3rd} & \textbf{4th} & \textbf{5th} \\  \hline
\textbf{Vasicek}                & 18.1         & 21.5         & 13.3         & 7.0          & 0.0          & 5.8          & 10.7          & 13.0         & 9.6         & 11.1          & 23.5         & 32.4         \\ 
\textbf{AFNS2}                  & 4.7          & 4.3          & 3.1          & 2.5          & 2.3          & 1.8          & 1.5          & 2.1          & 1.3         & 1.8          & 4.3          & 8.0          \\ 
\textbf{AFNS3}                  & 3.1          & 3.1          & 3.3          & 2.7          & 2.1          & 1.6          & 1.6          & 1.0          & 1.3         & 1.8          & 0.9          & 1.0          \\ 
\textbf{Shadow}                  & 2.9          & 3.1          & 3.3          & 2.6          & 2.1          & 1.6          & 1.6          & 0.9          & 1.3         & 1.8          & 0.9          & 1.8        \\   \hline
\end{tabular}
\caption{RMSEs of the fitted futures rates based on the final parameter estimates and filtered state variables for each of the models using the full SOFR data sample. All values are in basis points. The final log-likelihood values of the Vasicek, AFNS2, AFNS3 and shadow rate model respectively are 41826.1, 51050.6, 53247.6 and 53409.3}
\label{SOFR_infit}
\end{table}
When assessing the fit of the model implied futures rates to the observed futures rates we consider the root mean square error defined as
\begin{align}
RMSE(i)=\sqrt{\frac{\sum_{t=1}^T \left(f^{obs}_t(i) -f_t(i)\right)^2}{T}},
\end{align}
where $f^{obs}_t(i)$ denotes the observed futures rate at time $t$ for contract number $i$ and $f_t(i)$ is the corresponding model implied futures rate. 
Table \ref{SOFR_infit} contains the RMSEs of the fitted futures rates for each model. Clearly, the Vasicek model is not able to properly fit the term structure resulting in large deviations from the observed rates. The two-factor model is able to capture most of the variation in the cross section of futures rates, but does show increasing RMSEs the short and long end of the term structure of futures rates, while the three-factor model provides a close fit across all observed contracts.
\begin{figure}[hbt!]
\centering
\includegraphics[angle=0, scale=.8]{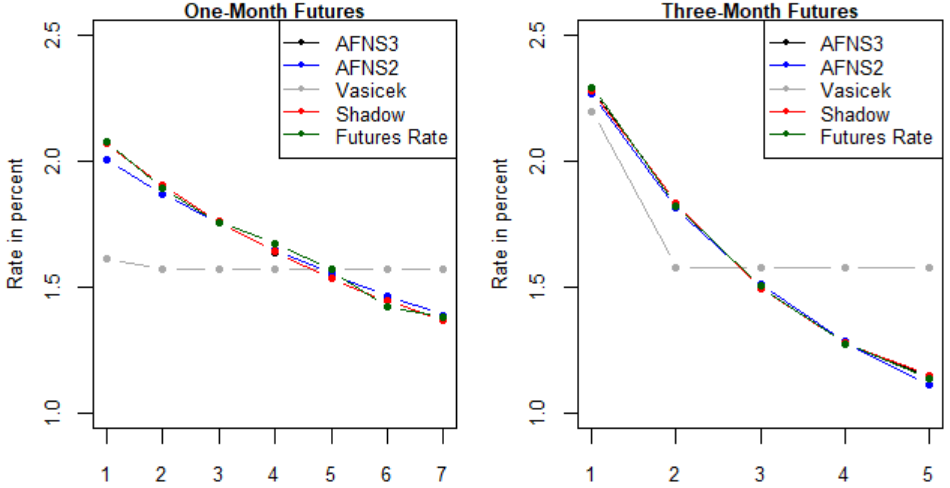}
\caption{Fitted and observed futures rates for the seven nearest one-month contracts and five nearest three-month contracts as of the 3rd of September 2019.}
\label{fig:fut_fit}
\end{figure}
This is also seen in figure \ref{fig:fut_fit}, which shows the fitted futures rates as of the 3rd of September 2019 together with the market implied futures rates. The futures rates on the particular date indicate a market expectation of a future decrease in SOFR. The graph illustrates the inability of the single-factor model in capturing the term structure of futures rates, while both the two- and three-factor AFNS models perform notably better. The nearest three-month futures rate appears to be reasonably well approximated by the Vasicek model. However, this is only due to the fact that the majority of the underlying overnight rates already have been accrued on this contract. The flat futures rate curve produced by the Vasicek model is a direct consequence of a near zero $K^Q$ estimate, making the state variable act as a level factor under the $Q$-dynamics. 
\begin{table}[hbt!]
\small
\center
\begin{tabular}{clllllllll}
\multicolumn{1}{l}{\textbf{Model}}                                                    & \textbf{Tenor} & \textbf{RMSE} & \textbf{Mean} & \textbf{SD} & \textbf{5\%} & \textbf{Q1} & \textbf{Median} & \textbf{Q3} & \textbf{95\%} \\ \hline
\textbf{Vasicek}                                                     & 1 Month        & 23.0          & -11.9         & 19.7                        & -54.0        & -26.5                 &  -0.5           & 2.7                  & 4.1           \\
                                                                                      & 3 Month        & 15.1          & -8.0         & 12.8                        & -34.1        & -18.3                 & -1.2          & 1.8                  & 2.9          \\
                                                                                      & 6 Month        & 6.7           & -3.3          & 5.8                         & -15.1        & -6.9                 & 0.1            & 0.8                 & 1.7           \\  \hline
\textbf{AFNS2}  & 1 Month        & 4.8           & 0.0          & 4.8                         & -12.4        & -0.7                  & 0.5             & 1.3                   & 4.9           \\
                                                                                      & 3 Month        & 1.9           & -0.2          & 1.9                         & -4.3         & -0.5                  & 0.1            & 0.6                   & 2.0           \\
                                                                                      & 6 Month        & 0.7           & 0.0          & 0.7                         & -0.9         & -0.2                  & 0.1            & 0.4                   & 1.1           \\  \hline
\textbf{AFNS3} & 1 Month        & 1.7           & 0.0           & 1.7                         & -2.8         & -0.7                  & -0.1             & 0.6                   & 3.1           \\
                                                                                      & 3 Month        & 1.0           & 0.1           & 0.9                         & -1.0         & -0.4                  & 0.0            & 0.4                   & 2.0           \\
                                                                                      & 6 Month        & 0.7           & 0.2          & 0.7                         & -0.6         & -0.3                  & 0.0            & 0.5                   & 1.4           \\  \hline
\textbf{Shadow} & 1 Month        & 1.8           & -0.3           & 1.8                         & -2.8         & -1.3                  & -0.5             & 0.5                   & 3.2           \\
                                                                                      & 3 Month        & 0.9           & 0.1           & 0.9                         & -1.3         & -0.5                  & 0.0            & 0.4                   & 1.9           \\
                                                                                      & 6 Month        & 0.7           & 0.1          & 0.7                         & -0.7         & -0.3                  & 0.0            & 0.4                   & 1.3         \\   \hline
\end{tabular}
\caption{Difference between model implied term rates based on rolling re-estimates and published Federal Reserve indicative term rates.}
\label{SOFR_tabel}
\end{table}
\\ \indent
Since a sufficiently liquid market of linear SOFR derivatives allowing for standard curve construction does not exist, we compare the model implied term rates to the model-independent term rates based on the method presented in \cite{heitfield2019inferring} published daily on the New York Federal Reserve's website\footnote{\url{https://www.federalreserve.gov/econres/notes/feds-notes/indicative-forward-looking-sofr-term-rates-20190419.htm}} as a benchmark reference. The model-free term rates in \cite{heitfield2019inferring} are calculated using a step function to parametrize the one-day forward curve, allowing for jumps on dates determined by the target rate announcement dates in the FOMC calender. Expected jump sizes are then recalibrated daily.
\\ \indent
Table \ref{SOFR_tabel} reports statistics on the differences between the indicative term rates published by the Fed and our model implied term rates. Comparing the root mean squared errors across the three models show that the three-factor model provides the closest fit to the model-independent term rates across all tenors. As also indicated by the graphs in figure \ref{fig:sofr} it is especially the shortest tenors that the lower factor models fail to accurately fit. \\The SOFR rate is heavily correlated with EFFR (See Figure \ref{fig:sofr_back}) and thus heavily influenced by the targets set by the FOMC, a reasonable assumption would be that a term structure model with a positive probability of jumps on the announcement dates would be necessary to fit the SOFR futures curve especially in the short term. Table \ref{SOFR_tabel} shows that this feature is in fact not necessary. The AFNS model, without introducing jumps, is able to produce term rates very similar to those of the model free step-function used in \cite{heitfield2019inferring}.
This is also seen in figure \ref{fig:sofr_comp} which compares the three-factor AFNS model three-month term rate to the indicative three-month term rate presented in \cite{heitfield2019inferring} as well as the realized forward-looking compounded SOFR three-month term rate.
\\ \indent
Figure \ref{fig:libor_comp} compares the three-month AFNS model implied SOFR term rate to the three-month-OIS rate and three-month LIBOR. The LIBOR-SOFR three-month term spread reflects a term premium present in the unsecured term borrowing of LIBOR.
Following the Great Financial Crisis the premium is driven by a credit risk and funding-liquidity risk premium (see e.g. \cite{backwell2019term}). 
The spread increases in times of stress as is clearly seen by the sharp increase in LIBOR following COVID-19 while SOFR, being an overnight secured rate, was not affected and like the OIS-rate closely tracked the target range set by the Federal Reserve.
Both the ARRC and ISDA recommend calculating the LIBOR-SOFR spread used for legacy contracts using a five-year historical median between the backward-looking compounded average SOFR and LIBOR. However, such a spread would not be consistent with a forward-looking term SOFR. Instead, adding a spread to a forward-looking term SOFR following the same approach would require historical SOFR term rates as presented in this paper. 

\subsection{Federal Funds Futures Implied Term Rates}
As a robustness check we re-estimate our model, using end of day data on federal funds futures traded at the CME instead of SOFR futures. Federal funds futures have traded since 1988 and thus allow us to test the models on a longer time frame. Also, a well developed OIS-market based on federal funds rates exists, which allows us to compare the model implied term rates to actual market term rates. We use data for one-, three- and six-month OIS contracts which have single end of term payment of exactly $\prod_{i=1}^{N} (1+d_i R_{d_i}^{EFFR}(t_i)) -1$, per unit notional with $R_{d_i}^{EFFR}$ being the daily quoted effective federal funds rate. From \eqref{prediction} it follows that the implied OIS term rate is exactly the EFFR equivalent of the term rate in \eqref{SOFRtermrate}.
\\ \indent
Federal Funds futures are only traded on a one-month contract period with listed contracts covering the 36 first calendar months. As with SOFR-futures most of the traded volume is concentrated around the nearest contracts. We therefore base the estimation on observed end of day prices on the seven nearest one-month futures contracts, thus covering term rates up to six months ahead. The data sample runs from January 2005 until December 2020 and we consider rolling daily estimates beginning in January 2007.
\begin{figure}[hbt!]
\centering
\includegraphics[angle=0, scale=0.63]{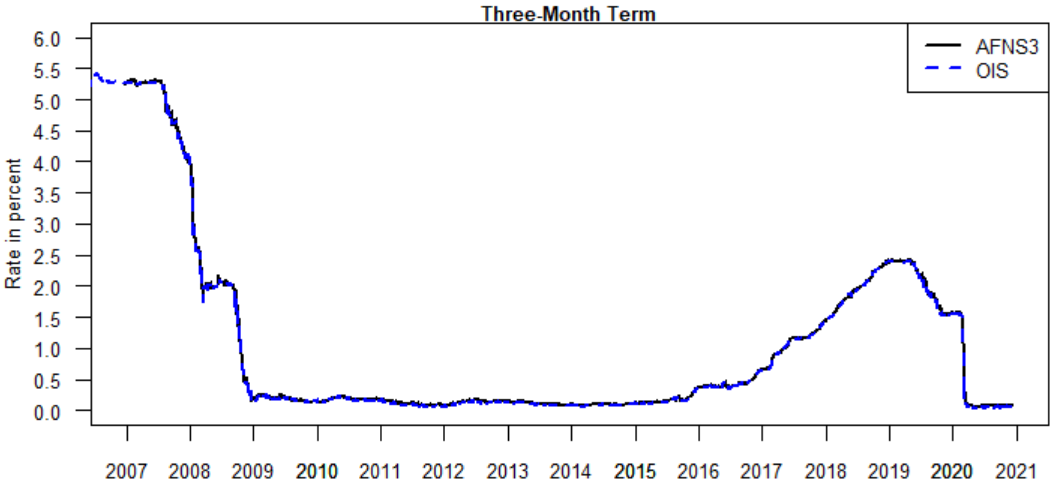}
\caption{Three-month end of day OIS-rate and model implied three-month three-month term rate based on rolling re-estimates from 3rd of January 2005.}
\label{fig:fed}
\end{figure}
\\ \indent
Figure \ref{fig:fed} plots the model implied term rates at a three month tenor against the corresponding end of day observed OIS-rate. Also in the federal funds case, the three-factor AFNS model is able to closely match the observed OIS-rates even during the drastic rate cuts seen under the financial crisis.
\begin{table}[hbt!]
\small
\center
\begin{tabular}{clllllllll}
\multicolumn{1}{l}{\textbf{Model}}                                                    & \textbf{Tenor} & \textbf{RMSE} & \textbf{Mean} & \textbf{SD} & \textbf{5\%} & \textbf{Q1} & \textbf{Median} & \textbf{Q3} & \textbf{95\%} \\ \hline
\textbf{Vasicek}                                                     & 1 Month        & 10.3          & -2.3          & 10.1                        & -22.5        & -2.7                  & -1.1             & 0.4                   & 9.9          \\
                                                                                      & 3 Month        & 4.0           & 0.0          & 4.0                         & -6.5        & -0.4                  & 0.5             & 1.3                   & 4.4           \\
                                                                                      & 6 Month        & 5.8           & 2.6          & 5.2                         & -4.5         & 0.2                  & 2.4 & 3.8                   & 11.0           \\  \hline
\textbf{AFNS2}  & 1 Month        & 2.5           & 0.2          & 2.5                         & -3.5         & -1.1                  & 0.5             & 1.4                   & 3.9           \\
                                                                                      & 3 Month        & 1.8           & 1.1           & 1.5                         & -1.1         & 0.5                  & 1.1 & 1.6                   & 3.1           \\
                                                                                      & 6 Month        & 1.7           & 1.2           & 1.2                         & -0.5         & 0.7                  & 1.2             & 1.6                   & 3.0           \\  \hline
\textbf{AFNS3} & 1 Month        & 2.7           & 1.3           & 2.4                         & -1.5         & 0.6                   & 1.2             & 1.8                   & 3.9           \\
                                                                                      & 3 Month        & 2.1           & 1.1           & 1.8                         & -1.2         & 0.4                   & 1.1             & 1.7                   & 3.4           \\
                                                                                      & 6 Month        & 1.8           & 1.2           & 1.3                         & -0.5         & 0.7                   & 1.2             & 1.6                   & 3.0           \\  \hline
\textbf{Shadow} & 1 Month        & 2.7           & 1.3           & 2.4                         & -1.5         & 0.5                   & 1.2             & 1.9                   & 3.9           \\
                                                                                      & 3 Month        & 2.1           & 1.1           & 1.9                         & -1.5         & 0.3                   & 1.1             & 1.7                   & 3.4           \\
                                                                                      & 6 Month        & 2.0           & 1.1           & 1.7                         & -1.8         & 0.6                   & 1.2             & 1.6                   & 3.1           \\ \hline 
\end{tabular}
\caption{Differences between model implied term rates based on rolling re-estimates and observed end of day OIS-rates.}
\label{term-ois}
\end{table}
\\ \indent
Table \ref{term-ois} reports statistics on the difference between model implied and observed OIS term rates. Interestingly, the results favour the two-factor model, which does provide the closest overall fit as measured by the RMSE, suggesting that two factors is enough to fit the EFFR market. This of course is partly due to the Federal Funds futures data only containing the seven first one-month contracts thus resulting in less daily data points that the model has to fit\footnote{A principal component analysis reveals that three factors are required to capture $99.9\%$ of the variation in the SOFR futures data, while only two factors are required for the EFFR futures data.}. We also note that the three-factor models slightly overestimate the term rates compared to the OIS rate, however still tracks the OIS implied term rate well. Refraining from the fact that this is based on end of day data from two different sources, the relatively lower liquidity of the short term OIS-market resulting in a higher spread compared to the futures market could have an impact on the slightly different term structures implied by these markets.
\subsection{Comparing SOFR and EFFR Futures}
\begin{figure}[hbt!]
\centering
\includegraphics[angle=0, scale=.55]{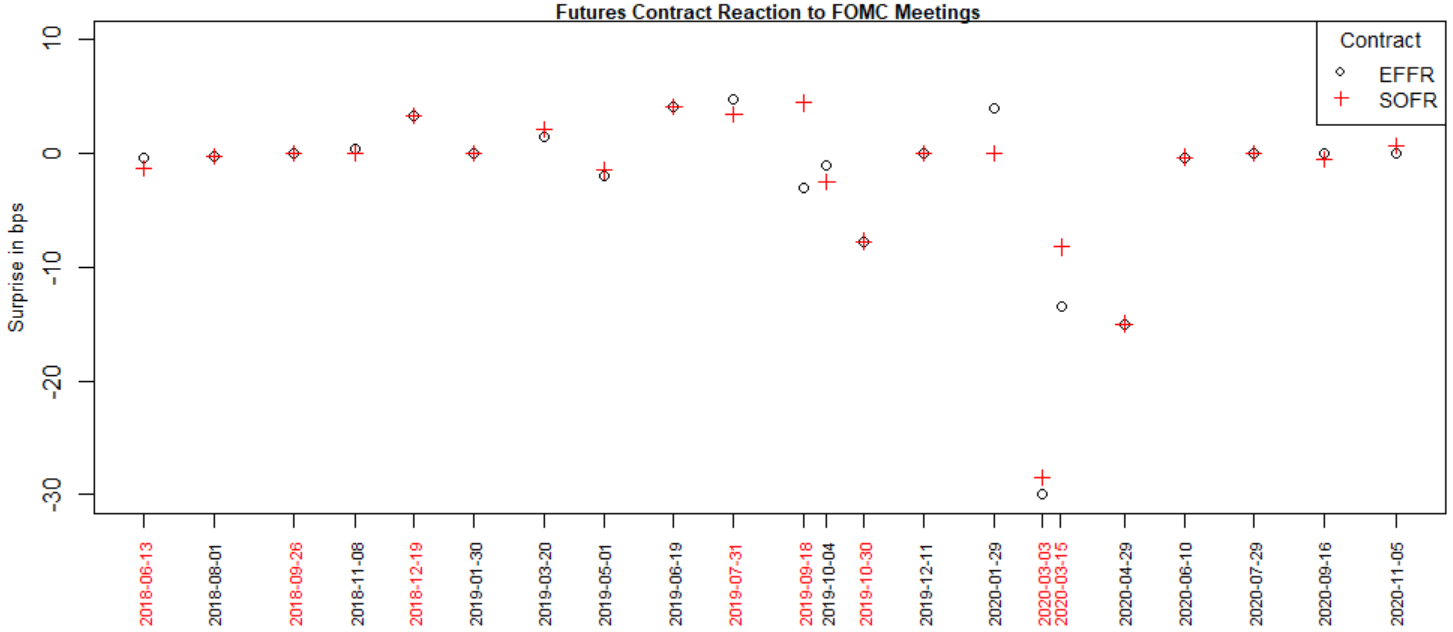}
\caption{Comparison of price changes on one-month spot EFFR and SOFR futures contracts on FOMC dates. Dates in red mark actual changes in the target rate.}
\label{fig:FOMC}
\end{figure}

As a final model-free robustness check we gauge the potential immaturity of the SOFR futures market by comparing the properties of the relatively new market for SOFR futures to the well-established Federal Funds based futures market. Since both the EFFR and SOFR are closely tied to the Federal Funds target range, we would expect that if the SOFR futures market is well functioning you would see similar response in both markets in the wake of the announcements following FOMC meetings. We follow the approach presented in \cite{kuttner2001monetary} to calculate the unexpected change in the Federal Funds target rate implied by the futures market. Let $\tau$ denote the day in the month of the FOMC meeting, $N$ the total number of days in the month and $f^{1m}_{\tau}$ the spot one-month futures rate the day of the FOMC meeting. The unexpected surprise in the futures rate, $\Delta r^u$, is then calculated as 
\begin{align}
    \Delta r^u = \frac{N}{N-\tau} (f^{1m}_{\tau}-f^{1m}_{\tau-1}).
\end{align}
When the meeting falls on the first date of the month, we replace $f^{1m}_{\tau-1}$ with the futures rate on the last day of the preceding month. If instead the meeting is at the last day of the month, we calculate the price difference in the futures contract of the following month.
Figure \ref{fig:FOMC} plots the unexpected change in the target rate implied by SOFR and EFFR futures on all FOMC dates since SOFR futures started trading on CME. 
The two markets react very similarly to FOMC meetings announcements. The largest difference of $7.5$ basis points appeared around the FOMC meeting date in September 2019. This is no surprise as SOFR experienced a drastic spike to above 5 percent on September 17 thus creating uncertainty around the SOFR and impacting the spot one-month futures rate.
\\ \indent
Next, we compare risk premia in the two markets following the method presented in \cite{piazzesi2008futures}. For comparability we focus solely on one-month futures contracts.
Using end of month data for the six nearest upcoming futures contracts, $n\in \{1,...,6\}$, we calculate the ex post realized excess returns of the one-month futures contract, $rx^{1m}_{t+n}$, as the difference between the time $t$ futures rate of the one-month futures contract, $f^{1m}_{(n)}(t)$, and the ex post realized one-month futures rates, $R^{1m}_{t+n}$,
\begin{align}
    rx^{1m}_{t+n}=f^{1m}_{(n)}(t)-R^{1m}_{t+n}.
\end{align}
We then proceed to run the regression
\begin{align}
    rx^{1m}_{t+n}=\alpha^{(n)}+\varepsilon_{t+n}^{(n)}.
\end{align}
The exercise is performed on both SOFR and EFFR futures based the same data sample starting with end of month futures data from May 2018 when SOFR futures launched on CME and until the end of December 2020. We also consider the full Federal Funds data sample used in the model estimation containing data from January 2005.
Table \ref{piazzesi1} contains the estimated constant risk premia for each of the expiries. We can first note that including the full data sample containing the post financial crisis years with little volatility in the Federal Funds overnight rate results in a much smaller risk premium than when using more recent data. But the table also demonstrates very similar risk-premia when comparing SOFR to EFFR futures. Interestingly, it shows a slightly higher premium in the SOFR contracts. One could speculate the reason could be due to liquidity risk premium in the rather new market for SOFR futures or perhaps more likely caused by the observed SOFR spikes. But as the table also shows, the statistical uncertainty is much too great to draw any inference in this regard. Most importantly, for this study, we do not find any evidence to suggest that SOFR futures should not be used in the construction of SOFR term rates. 
\begin{table}[hbt!]
\small
\noindent\makebox[\textwidth]{%
\begin{tabular}{ccccccc}
\textbf{n} & \multicolumn{2}{c}{\textbf{SOFR}}                                                                            & \multicolumn{2}{c}{\textbf{EFFR}}                                                                            & \multicolumn{2}{c}{\textbf{EFFR Full Sample}}                                                              \\ \hline
           & $\alpha^{(n)}$                                        & Annualized                                            & $\alpha^{(n)}$                                        & Annualized                                            & $\alpha^{(n)}$                                       & Annualized                                           \\ \hline
1m         & \begin{tabular}[c]{@{}c@{}}2.5\\ (2.5)\end{tabular}   & \begin{tabular}[c]{@{}c@{}}30.1\\ (30.0)\end{tabular} & \begin{tabular}[c]{@{}c@{}}2.3\\ (2.3)\end{tabular}   & \begin{tabular}[c]{@{}c@{}}27.3\\ (27.8)\end{tabular} & \begin{tabular}[c]{@{}c@{}}1.6\\ (0.6)\end{tabular}  & \begin{tabular}[c]{@{}c@{}}19.0\\ (7.4)\end{tabular} \\ \hline
2m         & \begin{tabular}[c]{@{}c@{}}7.4\\ (5.1)\end{tabular}   & \begin{tabular}[c]{@{}c@{}}44.5\\ (30.4)\end{tabular} & \begin{tabular}[c]{@{}c@{}}6.7\\ (4.9)\end{tabular}   & \begin{tabular}[c]{@{}c@{}}40.2\\ (29.5)\end{tabular} & \begin{tabular}[c]{@{}c@{}}4.3\\ (1.4)\end{tabular}  & \begin{tabular}[c]{@{}c@{}}25.8\\ (8.1)\end{tabular} \\ \hline
3m         & \begin{tabular}[c]{@{}c@{}}12.6\\ (6.9)\end{tabular}  & \begin{tabular}[c]{@{}c@{}}50.2\\ (27.7)\end{tabular} & \begin{tabular}[c]{@{}c@{}}11.8\\ (6.7)\end{tabular}  & \begin{tabular}[c]{@{}c@{}}47.1\\ (26.9)\end{tabular} & \begin{tabular}[c]{@{}c@{}}7.1\\ (2.0)\end{tabular}  & \begin{tabular}[c]{@{}c@{}}28.6\\ (7.9)\end{tabular} \\ \hline
4m         & \begin{tabular}[c]{@{}c@{}}18.5\\ (8.4)\end{tabular}  & \begin{tabular}[c]{@{}c@{}}55.4\\ (25.3)\end{tabular} & \begin{tabular}[c]{@{}c@{}}17.4\\ (8.3)\end{tabular}  & \begin{tabular}[c]{@{}c@{}}52.1\\ (24.8)\end{tabular} & \begin{tabular}[c]{@{}c@{}}10.1\\ (2.6)\end{tabular} & \begin{tabular}[c]{@{}c@{}}30.3\\ (7.7)\end{tabular} \\ \hline
5m         & \begin{tabular}[c]{@{}c@{}}25.4\\ (9.5)\end{tabular}  & \begin{tabular}[c]{@{}c@{}}61.0\\ (22.8)\end{tabular} & \begin{tabular}[c]{@{}c@{}}24.3\\ (9.3)\end{tabular}  & \begin{tabular}[c]{@{}c@{}}58.5\\ (22.3)\end{tabular} & \begin{tabular}[c]{@{}c@{}}13.2\\ (3.1)\end{tabular} & \begin{tabular}[c]{@{}c@{}}31.6\\ (7.4)\end{tabular} \\ \hline
6m         & \begin{tabular}[c]{@{}c@{}}33.3\\ (10.4)\end{tabular} & \begin{tabular}[c]{@{}c@{}}66.6\\ (20.9)\end{tabular} & \begin{tabular}[c]{@{}c@{}}32.5\\ (10.1)\end{tabular} & \begin{tabular}[c]{@{}c@{}}65.0\\ (20.2)\end{tabular} & \begin{tabular}[c]{@{}c@{}}16.3\\ (3.6)\end{tabular} & \begin{tabular}[c]{@{}c@{}}32.6\\ (7.2)\end{tabular} \\ \hline
\end{tabular}
}
\caption{Risk premium for one-month SOFR and EFFR futures contracts using futures data from end of May 2018 until end of December 2020 as well as the full Federal Funds futures sample starting in January 2005.  Standard errors are shown in parentheses. Annualized excess returns are obtained by multiplying the excess returns, $rx_{t+n}^{(n)}$, with $1/n$ before running the regression.}
\label{piazzesi1}
\end{table}

\subsection{SOFR Volatility}
\begin{figure}[h!]
\centering
\includegraphics[angle=0, scale=.7]{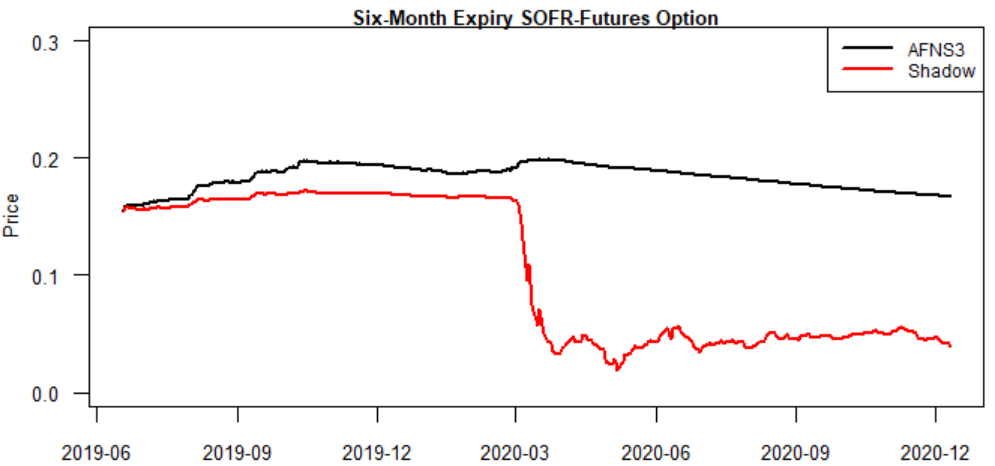}
\caption{The price of an at the money call option on a hypothetical three-month SOFR futures contract expiring in six months. The prices are quoted in IMM index points and based on daily rolling re-estimations and obtained using Monte Carlo simulations under the risk-neutral measure.}
\label{fig:sofr_option}
\end{figure}

The Gaussian short rate distribution of the AFNS model ignores the decrease in volatility resulting from rates being compressed at a lower bound. As of December 2020 the Federal Reserve has never lowered its target rate below zero or expressed willingness in doing so. Therefore, in this paper we assume a zero lower bound of U.S. rates as is often done in shadow rate models on U.S. data, see e.g. \cite{christensen2016modeling}. However, towards the end of 2020 Federal Funds futures contracts have priced have priced negative rates in 2021.
In the following section we show how the compression of rates at the zero lower bound also impacts the pricing of options on futures. In order to compare the change in value across time figure \ref{fig:sofr_option} plots the value of a hypothetical at the money call option on a three-month SOFR European futures option with a fixed six month expiry.\footnote{In reality, CME SOFR futures options are American style and thus have an early exercise premium. \cite{flesaker1993testing} reports that the the premium is typically less than one basis point for options that are not substantially in the money.} The option value can then be calculated as
\begin{align}
    C(t,P^{3m}(t;S,T),K,T)=\mathbb{E}^Q\left[e^{-\int_t^T r_s ds} \left(P^{3m}(T;S,T)-K\right)^+ |\mathcal{F}_t\right]
\end{align}
where $P^{3m}(t;S,T)$ denotes the time $t$ price of the three-month SOFR futures contract expiring at time $T$.
Figure \ref{fig:sofr_option} clearly shows a decrease in the value of the option in the shadow rate model as the distribution of future short rates becomes truncated at the zero lower bound. This is especially apparent after the rate drop in March 2020 where the option hit a minimum value equal to just $6\%$ of the price of the equivalent option in the AFNS model.

\section{SOFR Futures Convexity Adjustment}
\begin{figure}[hbt!]
\centering
\includegraphics[angle=0, scale=.7]{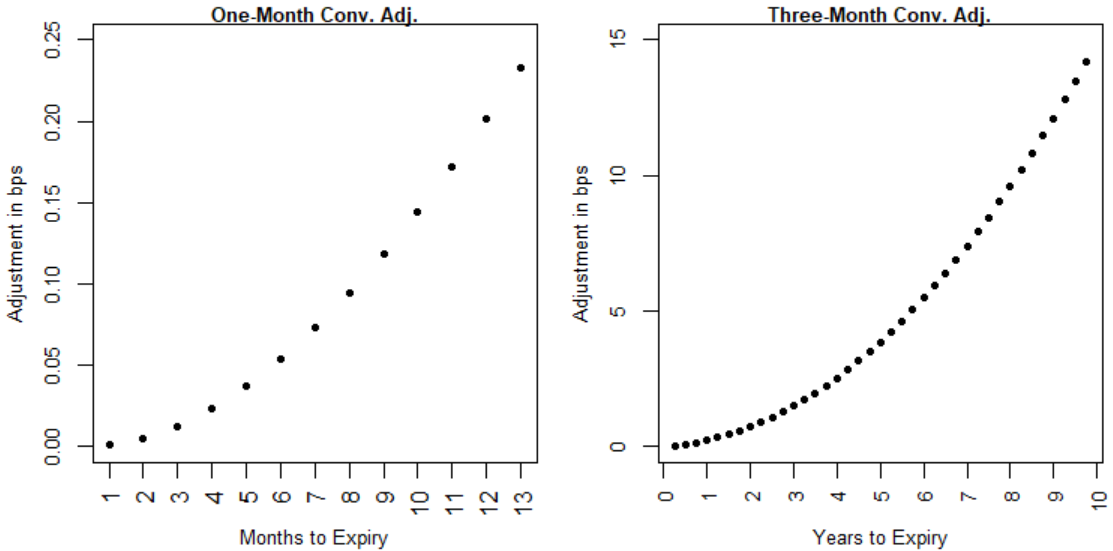}
\caption{Size of convexity adjustments in the AFNS model as of December 11th 2020 using the final set of SOFR-estimates based on the full sample. The adjustments are plotted for all open SOFR futures contracts traded on the CME.}
\label{fig:conv_adj}
\end{figure}

\begin{figure}[hbt!]
\centering
\includegraphics[angle=0, scale=.7]{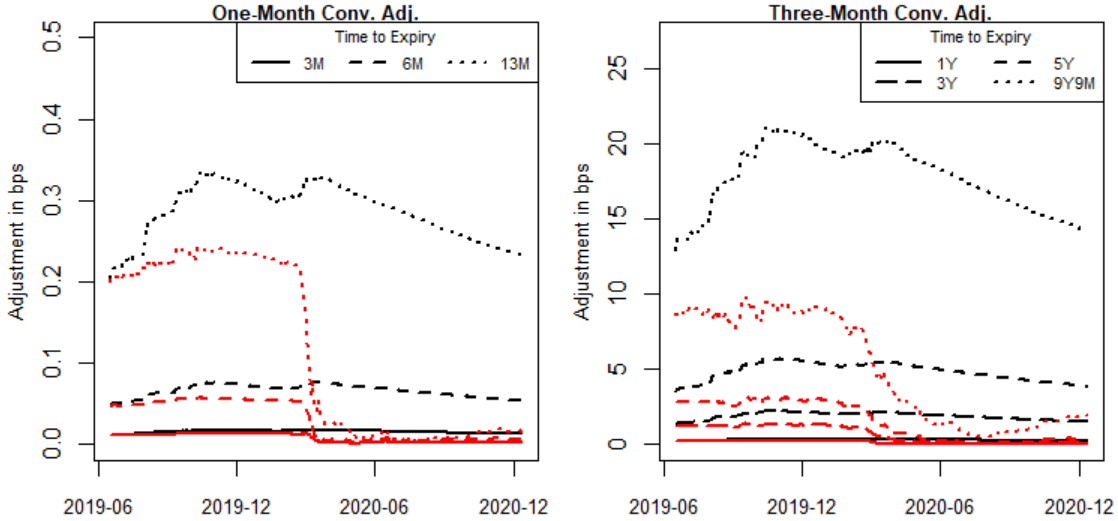}
\caption{Size of convexity adjustments using rolling re-estimates on SOFR futures data for different fixed time to expiries. The 13 months to expiry one-month and 9 years and 9 months to expiry three-month contracts reflect the last open SOFR futures contracts. The black lines plot the standard AFNS model and the red lines plot the equivalent convexity adjustment in the shadow rate extension.}
\label{fig:conv_adj2}
\end{figure}

\begin{figure}[hbt!]
\centering
\includegraphics[angle=0, scale=.7]{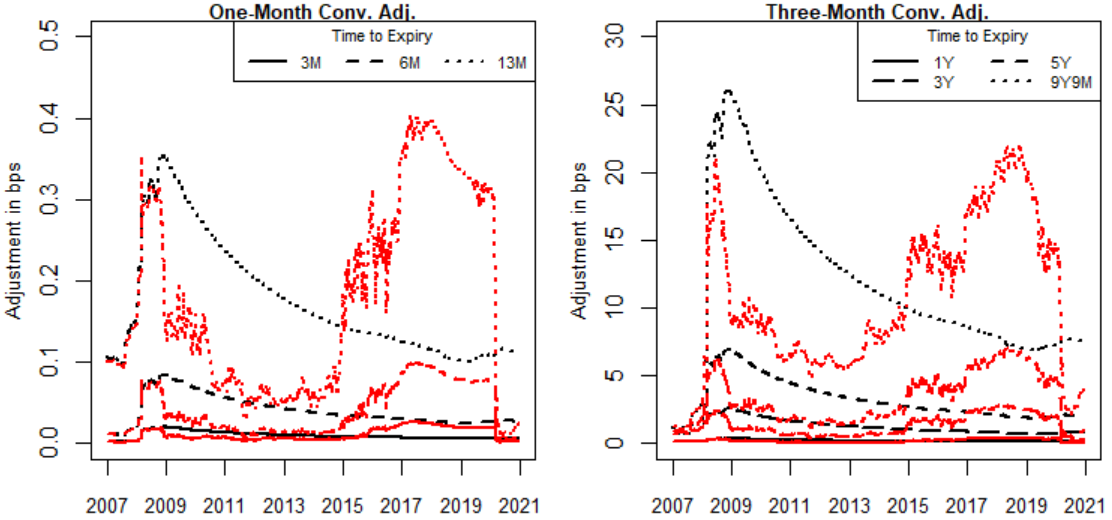}
\caption{Size of convexity adjustments using rolling re-estimates on federal funds futures data. The 13 months to expiry one-month and 9 years and 9 months to expiry three-month contracts reflect the last open SOFR futures contracts.}
\label{fig:conv_effr}
\end{figure}

The convexity adjustment of a futures contract is defined as the difference between the futures rate and an equivalent forward rate. Having estimated a model for the term structure of futures rates, we can compute the convexity adjustment. We follow \cite{mercurio2018simple} when defining the convexity adjustment for the one- and three-month SOFR futures. Here we focus solely on the three-factor AFNS model as well as its shadow rate extension.
The one-month adjustment is calculated from the continuously compounded forward rate, $F(t;S,T)$, given by
\begin{align}
F(t;S,T)=\frac{\log(p(t,S))-\log(p(t,T))}{T-S}.
\end{align}
Defining the convexity adjustment as the difference between the one-month futures rate and the continuously compounded forward rate, $C^{1m}(t;S,T):=f^{1m}(t;S,T)-F(t;S,T)$, and applying (\ref{1mfut_afns}) the one-month adjustment in the AFNS model is
\begin{align}
\label{1mconv}
C^{1m}(t;S,T)=\frac{A(t,T)-A(t,S)}{T-S}.
\end{align}
We define the three-month convexity adjustment as the difference between the three-month futures rate and the simple forward rate defined as,
\begin{align}
R^F(t;S,T)=\frac{1}{T-S}\left(\frac{p(t,S)}{p(t,T)}-1\right).
\end{align}
Similarly the convexity adjustment is defined as $C^{3m}(t;S,T):=f^{3m}(t;S,T)-R^F(t;S,T)$. Then by \eqref{3mfut_afns} in the appendix, the three-month convexity adjustment in the AFNS model is
\begin{align}
\label{3mconv}
C^{3m}(t;S,T)=\left(R^F(t;S,T)+\frac{1}{T-S}\right)\left(e^{A(t,T)-A(t,S)}e^{A(S,T)}e^{\frac{1}{2}B(S,T)'\mathbb{V}^{Q}[X_S|\mathcal{F}_t]B(S,T)}-1 \right).
\end{align}
One- and Three-month convexity adjustments in the shadow rate model are calculated using Monte Carlo simulations.
Figure \ref{fig:conv_adj} plots the AFNS model implied size of the convexity adjustment of the futures rates for the 13 and 39 one- and three-month contracts matching the set of listed SOFR futures on the CME, assuming one is at the beginning of the accrual period of the nearest futures. It clearly shows that using long-dated three-month futures rates as a proxy for the forward rate without any adjustment can lead to sizeable errors. However, when focusing solely on the nearest futures contracts covering the short end of the term structure the convexity adjustment is of very limited scale.
\\ \indent
Since the convexity adjustment is model dependent, it changes with the model estimates. To illustrate the range of the size of the adjustment for our data period, we plot convexity adjustments for a selected set of contracts based on rolling re-estimates using SOFR futures data. This is illustrated in figure \ref{fig:conv_adj2}, which shows a notable variation in the size of the convexity adjustment for long-dated three-month futures rates.
\\ \indent
Given the expression for the one-month convexity adjustment (\ref{1mconv}), we see that it is dependent on the estimates of the decay parameter, $\lambda$, and the volatility matrix, $\Sigma$. As seen in (\ref{3mconv}) the three-month convexity adjustment also depends on the level of the discrete forward rate $R^F(t;S,T)$, however, only to a limited extent since the term is small compared to $\frac{1}{T-S}\approx 4$. Given the final set of parameter values increasing the forward rate by $10\%$ adds less than $0.5$ basis point to the convexity adjustment of the furthest futures contract. The main factor driving the variation in the convexity adjustment is the conditional yield volatility under the risk neutral measure as represented by the term $B(S,T)'\mathbb{V}^{Q}[X_S|\mathcal{F}_t]B(S,T)$ in the three-month convexity adjustment. Thus when rates become increasingly volatile the size of the convexity adjustment also increases.
Figure \ref{fig:conv_adj2} also plots the equivalent convexity adjustment of the shadow rate model. The plots show the convexity adjustment of all contracts dropping close to zero after the FED lowered the target rate in March 2020. The drop reflects a decrease in volatility in the shadow rate model and highlights that the convexity adjustment is zero when rates are deterministic.
\\ \indent
To obtain estimates of the convexity adjustments for a longer sample period containing the global financial crisis, we consider the estimates from the federal funds futures data and plot convexity adjustments as if both one- and three-months contracts equal to those of the SOFR market traded in this market. Figure \ref{fig:conv_effr} clearly illustrates a spike in the size of the adjustment following the crisis. Subsequently, when rates became compressed against the zero lower bound the convexity adjustment gradually decreases in the Gaussian AFNS model, while in the shadow rate model we again see a significant drop.
Due to the short rate specification in the shadow rate model, the volatility parameters do not need to decrease to fit the compression in volatility while at the lower bound,
thus when the Federal Funds increased the Federal Funds target range away from zero in 2015 the convexity adjustment increases noticeably.

\section{Conclusion}
In this paper we investigate the use of dynamic term structure models for the SOFR futures market using historical data. We find that a standard three-factor arbitrage-free Nelson-Siegel model is able to describe the dynamics of futures prices quite well without explicitly incorporating the expected jumps in the spot rate on FOMC announcement dates. Furthermore, our model is able to produce term rates that are indistinguishable from the indicative rates published by the federal reserve. We furthermore demonstrate that how a shadow rate extension can explicitly capture the volatility compression occruring after a rate drop.  Using the models we are able to quantify the size of the convexity correction in this market, and determine at what maturity and volatility level such a correction should be implemented. 

This provides a framework that can easily be made to include other derivatives such as SOFR caps, floors and swaptions as these markets become more mature. Future work could extend the empirical analysis to a multi-curve setup. This could include estimating the joint behavior of multiple benchmarks suchs as SOFR, EFFR and LIBOR. The estimated models could also be directly used as the SOFR or EFFR components to one of the many existing multi-curve setups (see for example \cite{grbac2015interest} for an overview) for pricing and risk managing derivatives on multiple benchmarks.

\section*{Acknowledgments}
The authors are grateful to Andrea Macrina for suggestions and comments.
\section*{Data Availability Statement}
The SOFR and EFFR futures data used in this paper were obtained through Refinitiv Eikon and used under license. The Indicative term rates published by the by the Federal Reserve as well as historical overnight SOFR and EFFR are available at the New York Federal Reserve's website.

\bibliography{sample}

\section{Appendix A: AFNS Model Specifications and Results} \label{afns3results}
Here we present the specific parameter restrictions of the AFNS models as well as results needed to price the one- and three-month futures contracts for the AFNS model, similar results can be obtained for the one- and two-factor models.
The three-factor Arbitrage-Free Nelson-Siegel model places the following restrictions on the risk neutral dynamics
\begin{align}
\theta^Q =\begin{pmatrix}
0 \\ 0 \\ 0
\end{pmatrix}, \quad
K^Q=
  \begin{pmatrix}
    0 & 0 &  0 \\
    0 & \lambda  & -\lambda \\
    0 & 0 & \lambda \\
  \end{pmatrix}, \quad
  \rho_0=0, \quad \rho_1 =\begin{pmatrix}
1 \\ 1 \\ 0
\end{pmatrix}.
\end{align}
The state variables are Gaussian with mean and variance
\begin{align}
\mathbb{E}^{Q}\left[X_T|\mathcal{F}_t \right] &=e^{-K^Q (T-t)} X_t,
\\
\mathbb{V}^{Q}\left[X_T|\mathcal{F}_t \right] &=
\int_t^T e^{-K^Q(T-s)}\Sigma \Sigma' \left(e^{-K^Q(T-s)}\right)' ds.
\end{align}
Where the matrix exponential can be calculated analytically as
\begin{align}
e^{-K^Q (T-t)} = \small
  \begin{pmatrix}
    1 & 0 &  0 \\
    0 & e^{-\lambda (T-t)}  & \lambda (T-t)e^{-\lambda (T-t)} \\
    0 & 0 & e^{-\lambda (T-t)} \\
  \end{pmatrix}.
\end{align}
Given a diagonal volatility matrix, $\Sigma$, we find
\begin{align*}
\label{afnsvar}
&\mathbb{V}^{Q}\left[X_T|\mathcal{F}_t \right]\\&= 
  \small \begin{pmatrix}
    \sigma_{11}^2 (T-t) & 0 &  0 \\
    0  &  \frac{-e^{-2  \lambda  (T-t)} + 1}{2\lambda} \sigma_{22}^2
	+ \frac{1-e^{-2  \lambda  (T-t)}\left((2  \lambda  (T-t))(\lambda  (T-t)+1 ) + 1\right)}{4\lambda} \sigma_{33}^2    
     &\frac{e^{-2  \lambda  (T-t)}(-2  \lambda  (T-t) - 1 ) + 1}{4\lambda} \sigma_{33}^2 \\
    0 & \frac{e^{-2  \lambda  (T-t)}(-2  \lambda  (T-t) - 1 ) + 1}{4\lambda} \sigma_{33}^2 & \frac{-e^{-2  \lambda  (T-t)} + 1}{2\lambda} \sigma_{33}^2 \\
  \end{pmatrix}.
\end{align*}
The integral $\int_t^T r_s ds$ is also Gaussian with mean and variance
\begin{align}
\mathbb{E}^Q\left[\int_t^T r_s ds |\mathcal{F}_t \right]&=-B(t,T)' X_t,\\
\mathbb{V}^Q\left[\int_t^T r_s ds |\mathcal{F}_t \right]&=\int_t^T \sum_{j=1}^3 \left(\Sigma' B(s,T) B(s,T)' \Sigma\right)_{jj} ds.
\end{align}
Where $B(t,T)$ takes the following form
\begin{align}
B(t,T)=
\begin{pmatrix}
-(T-t) \\ -\frac{1-e^{-\lambda(T-t)}}{\lambda} \\ (T-t)e^{-\lambda(T-t)}-\frac{1-e^{-\lambda(T-t)}}{\lambda}
\end{pmatrix}.
\end{align}
Define $A(t,T):=\frac{1}{2}\mathbb{V}^Q\left[\int_t^T r_s ds |\mathcal{F}_t \right]$, an analytical formula of $A(t,T)$ is calculated in the original paper \cite{christensen2011affine}.
Assuming a diagonal volatility matrix the expression simplifies to
\begin{align*} 
&A(t,T)
 = \sigma_{11}^2  \frac{(T-t)^3}{6}
+\sigma_{22}^2  (T-t)
\left(\frac{1}{2\lambda^2}-\frac{1}{\lambda^3}\frac{1-e^{-\lambda(T-t)}}{T-t}+\frac{1}{4\lambda^3}\frac{1-e^{-2\lambda(T-t)}}{T-t} \right) \\
&+\sigma_{33}^2  (T-t)
\left(\frac{1}{2\lambda^2}+\frac{1}{\lambda^2}e^{-\lambda(T-t)}-\frac{1}{4\lambda}(T-t)e^{-2\lambda(T-t)}-\frac{3}{4\lambda^2}e^{-2\lambda(T-t)} - \frac{2}{\lambda^3} \frac{1-e^{-\lambda(T-t)}}{T-t} + \frac{5}{8\lambda^3}\frac{1-e^{-2\lambda(T-t)}}{T-t} \right).
\end{align*}
We are now ready to compute the one- and three-month futures rates. The one-month futures rate can be computed as
\begin{align}
\label{1mfut_afns}
f^{1m}(t;S,T)&=\mathbb{E}^Q\left[\frac{1}{T-S}\int_S^T r_s ds|\mathcal{F}_t \right]  \nonumber\\
&=\mathbb{E}^Q\left[\frac{1}{T-S}\int_t^T r_s ds -\frac{1}{T-S}\int_t^S r_s ds \mathcal{F}_t\right]  \nonumber\\
&=\frac{\left(B(t,S)-B(t,T)\right)'}{T-S}X_t.
\end{align}
And when $S<t$ such that part of the underlying has already been realized
\begin{align}
f^{1m}(t;S,T)=\frac{1}{N} \sum_{i=1}^{N_0} R_{d_i}(t_i)-\frac{B(t,T)'}{T-S}X_t.
\end{align}
The three-month futures rate can be computed using the tower property
\begin{align}
\mathbb{E}^Q\left[e^{\int_S^T r_s ds}| \mathcal{F}_t\right]
&=\mathbb{E}^Q\left[\mathbb{E}^Q\left[e^{\int_S^T r_s ds}  |\mathcal{F}_S \right]|\mathcal{F}_t \right] \nonumber\\
&=\mathbb{E}^Q\left[e^{A(S,T)-B(S,T)'X_S}|\mathcal{F}_t \right]\nonumber\\
&=e^{A(S,T)}e^{-B(S,T)'e^{-K^Q (S-t)} X_t+\frac{1}{2}B(S,T)'\mathbb{V}^{Q}[X_S|\mathcal{F}_t]B(S,T)}.
\end{align}
such that
\begin{align*}
\label{3mfut_afns}
f^{3m}(t;S,T)&=\frac{1}{T-S}\left(e^{A(S,T)}e^{-B(S,T)'e^{-K^Q (S-t)} X_0+\frac{1}{2}B(S,T)'\mathbb{V}^{Q}[X_S|\mathcal{F}_t]B(S,T)}-1 \right).
\end{align*}
And when $S<t$
\begin{align}
f^{3m}(t,S,T)
=\frac{1}{T-S}\left(\left(\prod_{i=1}^{N_0} \left[1+d_i R_{d_i}(t_i) \right]\right)e^{A(t,T)-B(t,T)'X_t}-1\right).
\end{align}

\section{Appendix B: Pricing the Futures Contract in a Shadow Rate Model}
Introducing a lower bound on the short rate results in a model where the short rate is no longer Gaussian. Therefore, we can not use the same results used to price the futures contracts in the standard AFNS model.
In this section we present formulas necessary to price the one- and three-month futures contracts in a shadow rate extension of the three-factor AFNS model.
We start by listing a few auxiliary results required to calculate the futures rates.
Using the results from section \ref{afns3results} we can calculate the mean of the shadow short rate as
\begin{align}
\mathbb{E}^{Q}[s_T |\mathcal{F}_t]&=\rho_1' e^{-K^Q (T-t)} X_t \nonumber \\
&=L_t+e^{-\lambda (T-t)}S_t+\lambda(T-t)e^{-\lambda (T-t)}C_t.
\end{align}
The variance is given by
\begin{align}
\mathbb{V}^{Q}[s_T |\mathcal{F}_t] 
= \rho_1' \mathbb{V}^{Q}[X_T |\mathcal{F}_t] \rho_1
\end{align}
and thus 
\iffalse
The variance is given by
\begin{align}
\mathbb{V}^{Q}[s_T |\mathcal{F}_t] =
\int_t^T \rho_1' e^{-K^Q(T-u)}\Sigma \Sigma' \left(e^{-K^Q(T-u)}\right)' \rho_1du.
\end{align}
The key object required to obtain expressions for the mean and variance is the matrix exponential of $-K^Q (T-t)$. Under the $K^Q$ specification of the AFNS model, The matrix exponential is given by
\begin{align}
e^{-K^Q (T-t)} = 
  \begin{pmatrix}
    1 & 0 &  0 \\
    0 & \exp(-\lambda(T-t))  & \lambda(T-t)\exp(-\lambda (T-t)) \\
    0 & 0 & \exp(-\lambda(T-t)) \\
  \end{pmatrix}.
\end{align}
Doing the necessary matrix multiplications and collecting terms we arrive at the integral
\ref{afnsvar}
\begin{align}
\mathbb{V}^{Q}[s_T |\mathcal{F}_t] =
\int_t^T  \sigma_{11}^2 + 
\sigma_{22}^2 e^{-2\lambda (T-u)}
+ \sigma_{33}^2 \left(\lambda^2 (T-u)^2 e^{-2\lambda(T-u)} \right)du.
\end{align}
Computing the integral we find
\fi
\begin{align}
\mathbb{V}^{Q}[s_T |\mathcal{F}_t] 
=\sigma_{11}^2(T-t) 
+ \sigma_{22}^2\frac{1-e^{-2\lambda(T-t)}}{2\lambda}+\sigma_{33}^2
\left(\frac{1-e^{-2\lambda(T-t)}}{4\lambda} - \frac{1}{2}(T-t)e^{-2\lambda (T-t)}- \frac{1}{2}(T-t)^2\lambda  e^{-2\lambda (T-t)}\right).
\end{align}
Where we have assumed a diagonal volatility matrix $\Sigma$ and used equation (\ref{afnsvar}). Furthermore, we note that for $s,u>t$
\begin{align}
Cov^{Q}[s_u,s_s |\mathcal{F}_t] =
\int_t^{u\wedge s} \rho_1' e^{-K^Q(u-\nu)}\Sigma \Sigma' \left(e^{-K^Q(s-\nu)}\right)' \rho_1 d\nu.
\end{align}
Again, assuming a diagonal volatility matrix the covariance can be calculated as
\begin{align}
Cov^{Q}[s_u,s_s |\mathcal{F}_t] 
&=\sigma_{11}^2( (u\wedge s)-t) 
+ \frac{\sigma_{22}^2}{2\lambda} \left(e^{-\lambda(s+u-2(u\wedge s))}
-e^{-\lambda(s+u-2t)} \right)  \nonumber \\
&+\frac{\sigma_{33}^2}{4\lambda}
\left(e^{-\lambda(s+u-2(u\wedge s))}\left(\lambda(s+u-2(u\wedge s))+1\right)
-e^{-\lambda(s+u-2t)}\left(2\lambda^2 (s-t)(u-t)+\lambda(s+u-2t)+1\right)\right).
\end{align}
The one- and three-month futures rates are computed using results from \cite{priebsch2013computing}. We follow the notation and denote $\mu_{t,T}=\mathbb{E}^{Q}[s_T |\mathcal{F}_t]$, $\sigma^2_{t,T}=\mathbb{V}^{Q}[s_T |\mathcal{F}_t]$ and $\sigma_{t,u\times s}=Cov^{Q}[s_u,s_s |\mathcal{F}_t]$
Computing the one-month futures rate requires calculating the first moment of $\int_t^T r_s ds$ under the risk neutral measure. The expectation is given by
\begin{align}
\label{1mshadow}
    \mathbb{E}^Q\left[\int_t^T r_s ds |\mathcal{F}_t \right]
    =\int_t^T 
    \mu_{t,s}
    \Phi \left(\frac{\mu_{t,s}}{\sigma_{t,s}} \right)
    \sigma_{t,s}
    \phi \left(\frac{\sigma_{t,s}}{\mu_{t,s}} \right)
    ds.
\end{align}
The one-month rate can therefore be computed up to the standard normal cdf, which is then numerically integrated over the remaining duration of the contract. \\
The three-month futures rate is approximated using the power series expansion of the cumulant generating function
\begin{align}
    \log \mathbb{E}^Q
    \left[e^{\int_t^T r_s ds} |\mathcal{F}_t \right]
    =\sum_{j=1}^\infty \frac{\kappa^Q_j}{j!}
\end{align}
where $\kappa^Q_j$ is the $j^{th}$ cumulant of $\int_t^T r_s ds$ under the $Q$-measure. We truncate the series at two terms and use the result that the two first cumulants of any random variable are equal to the two first centered moments.
The approximation then becomes
\begin{align}
    \mathbb{E}^Q
    \left[e^{\int_t^T r_s ds} |\mathcal{F}_t \right]
    \approx\exp\left(\mathbb{E}^Q\left[\int_t^T r_s ds |\mathcal{F}_t \right]+
    \frac{1}{2}\mathbb{V}^Q\left[\int_t^T r_s ds |\mathcal{F}_t \right]\right).
\end{align}
Since we have already shown how to compute the first moment in the one-month futures rate, it only remains to compute the
second moment to obtain the variance. Again, using results from \cite{priebsch2013computing} the expression for the second moment under the risk neutral measure is
\begin{align}
    \mathbb{E}^Q\left[\left(\int_t^T r_s ds\right)^2 |\mathcal{F}_t \right]
    =&2\int_t^T \int_t^s \Bigg\lbrace
     \left(\mu_{t,u}\mu_{t,s}+\sigma_{t,u\times s}\right)
     \Phi_2^d\left(-\zeta_{t,u},-\zeta_{t,s};\chi_{t,u\times s}\right)  \nonumber\\
     &+\sigma_{t,s}\mu_{t,u}\phi(\zeta_{t,s})
     \Phi \left(\frac{\zeta_{t,u}-\chi_{t,u\times s}\zeta_{t,s}}{\sqrt{1-\chi_{t,u\times s}^2}} \right) \nonumber \\
     &+\sigma_{t,u}\mu_{t,s}\phi(\zeta_{t,u})
     \Phi \left(\frac{\zeta_{t,s}-\chi_{t,u\times s}\zeta_{t,u}}{\sqrt{1-\chi_{t,u\times s}^2}} \right) \nonumber \\
     &+\sigma_{t,u}\sigma_{t,s}
     \sqrt{\frac{1-\chi_{t,u\times s}^2}{2\pi}}
     \phi\left(\sqrt{\frac{\zeta_{t,u}^2-2*\chi_{t,u\times s}\zeta_{t,u}\zeta_{t,s}+\zeta_{t,s}^2}{1-\chi_{t,u\times s}^2}} \right) \Bigg\rbrace 
    duds
\end{align}
with $\zeta_{t,j}=\frac{\mu_{t,j}}{\sigma_{t,j}}$ for $j=u,s $ and $\chi_{t,u\times s}=\frac{\sigma_{t,u\times s}}{\sigma_{t,u}\sigma_{t,s}}$. $\Phi_2^d$ denotes the decumulative bivariate normal cdf which satisfies $\Phi_2^d(z_1,z_2;\chi)=1-\Phi(z_1)-\Phi(z_2)+\Phi_2(z_1,z_2;\chi)$, where $\Phi_2$ is the cumulative bivariate normal distribution function of two standard normals with covariance $\chi$. As with the first moment the integrand is numerically integrated over both dimensions to obtain the second moment \footnote{Computing the double integral can quickly become computationally expensive. In our empirical implementaion we numerically integrate the expression using the Gauss-Kronrod method with five points in each dimension.}. \\
Zero coupon bonds in the shadow rate model are approximated using the series expansion of the cumulant generating function
\begin{align}
    \log \mathbb{E}^Q
    \left[e^{-\int_t^T r_s ds} |\mathcal{F}_t \right]
    =\sum_{j=1}^\infty (-1)^j \frac{\kappa^Q_j}{j!}.
\end{align}
Again, truncating the series at two terms the approximation becomes
\begin{align}
    P(t,T)=
    \mathbb{E}^Q
    \left[e^{-\int_t^T r_s ds} |\mathcal{F}_t \right]
    =\exp\left(-\mathbb{E}^Q\left[\int_t^T r_s ds |\mathcal{F}_t \right]+
    \frac{1}{2}\mathbb{V}^Q\left[\int_t^T r_s ds |\mathcal{F}_t \right]\right).
\end{align}
\subsection{Accuracy of The Shadow Model Futures Rate Approximation}
We test the accuracy of the cumulant based approximation of the futures rates in the shadow rate model to the futures rates in equation \ref{1mapp} and \ref{3mapp}. To gauge the size of the error we compare the approximation to futures rates obtained using Monte Carlo simulations under the risk neutral measure. The simulations are based on 100,000 paths with a step size of $1/3600$.
Table \ref{Shadow_accuracy} contains the error as measured by the difference between the approximation and Monte Carlo value for all maturities used in the estimation. We assume that the prices are observed at the beginning of the nearest futures contract and thus none of the overnight rates have been accrued by the contracts. To evaluate the approximations both away from and at the zero lower bound we consider the estimates and state variables at the beginning of the estimation period as well as the most recent date reflecting a period with rates at their lower bound.

\begin{table}[hbt!]
\small
\center
\begin{tabular}{ccccccccccccc}
\textbf{Date} & \multicolumn{7}{c}{\textbf{One-Month}}                                                                 & \multicolumn{5}{c}{\textbf{Three-Month}}                                \\ \hline
                                & \textbf{1M} & \textbf{2M} & \textbf{3M} & \textbf{4M} & \textbf{5M} & \textbf{6M} & \textbf{7M} & \textbf{3M} & \textbf{6M} & \textbf{9M} & \textbf{12M} & \textbf{15M} \\ \hline
\textbf{06/17/19}                & -0.02         & -0.02         & -0.01         & -0.01          & -0.01          & 0.00          & 0.00          & 0.14         & 0.07         & 0.04          & 0.02         & 0.00         \\
\textbf{12/11/20}                  & 0.00          & 0.04          & 0.02          & 0.06          & 0.07          & 0.04          & 0.03          & 0.02          & 0.06          & 0.02          & -0.02          & -0.04          \\ \hline
\end{tabular}
\caption{Difference between the approximated futures rate and Monte Carlo implied futures rate for all contract lengths used in the estimation. All values are in basis points.}
\label{Shadow_accuracy}
\end{table}

Table \ref{Shadow_accuracy} illustrates that errors on all considered contracts are a fraction of a basis point and far less than the minimum price fluctuation on CME SOFR futures contracts. Thus for estimation purposes the approximation results in accurate futures prices in the shadow rate model. 

\section{Appendix C: Estimation Methodology}
The presented models are estimated using a Kalman Filter on end of day futures data.
Thus we consider the discretized state process under the $P$-dynamics given by
\begin{align}
X_{t}=\theta^P+e^{-K^P \Delta t} (X_{t-1}-\theta^P) + \int_0^{\Delta t}  e^{-K^P u}\Sigma dW_u^{P}
\end{align}
with $\Delta t$ set to $1/250$ to approximately reflect the number of daily futures data observations in a year.
Rearranging we define the state equation of the Kalman Filter as
\begin{align}
X_{t}=\left( I-e^{-K^P \Delta t}\right)\theta^P + e^{-K^P \Delta t} X_{t-1} + \xi_t
\end{align}
with $\xi_t \sim \mathcal{N}(0,Q)$ and covariance, Q, where
\begin{align}
Q=\int_0^{\Delta t}  e^{-K^P u}\Sigma \Sigma' \left(e^{-K^P u}\right)'du.
\end{align}
For any specification of $K^P$ an analytical solution to Q is available by applying the eigendecomposition with $V$ a matrix of the eigenvectors and $\Lambda$ a vector of the corresponding eigenvalues
\begin{align}
e^{-K^P u}=e^{ -V \Lambda V^{-1} u} =Ve^{ - \Lambda  u}V^{-1},
\end{align}
we can calculate the covariance matrix as
\begin{align}
Q =V \left(\int_0^{\Delta t}  e^{- \Lambda  u} V^{-1}\Sigma \Sigma' 
(V^{-1}) ' e^{ - \Lambda  u} du \right) V' .
\end{align}
Define $G=V^{-1}\Sigma \Sigma' (V^{-1}) '$ the entries of the matrix defined by the integral are then
\begin{align}
&\left(\int_0^{\Delta t}  e^{ -\Lambda  u} V^{-1}\Sigma \Sigma' (V^{-1}) ' e^{ - \Lambda  u} du \right)_{ij}
=\frac{G_{ij}}{\Lambda_{ii}+\Lambda{jj}} \left(1 - e^{ -(\Lambda_{ii}+\Lambda{jj})\Delta t }\right).
\end{align}
The prediction step in the Kalman filter is computed as
\begin{align}
&X_{t|t-1}=F_t X_{t-1|t-1} + C_t, \\
&P_{t|t-1}=F_t P_{t-1|t-1} F_t' + Q.
\end{align}
Where $F_t=e^{-K^P \Delta t} $ denotes the state transition model and $C_t=\left( I-e^{-K^P \Delta t}\right)\theta^P$ the control input. The standard Kalman filter requires the measurement equation to be affine in the state vector
\begin{align}
y_t = A_t+B_t X_t+\varepsilon_t
\end{align}
with $\varepsilon_t \sim \mathcal{N}(0,H)$ where $H$ is a diagonal matrix and the transition and measurement errors, $\xi_t$ and $\varepsilon_t$, are independent.
The affine measurement equation is only satisfied for the one-month futures rate approximation. in the case of the three-month futures we apply the method of the extended Kalman filter where the measurement equation is assumed to be on the general form
\begin{align}
y_t=h(X_t,\Theta)+\varepsilon_t.
\end{align}
A Taylor expansion of $h$ is then done around $X_{t|t-1}$
\begin{align}
h(X_t,\Theta)\approx h(X_{t|t-1},\Theta)+\frac{\partial h(X_t,\Theta)}{\partial X_t}\at[\Big]{X_t=X_{t|t-1}} (X_t-X_{t|t-1}).
\end{align}
Now defining
\begin{align}
&A_t= h(X_{t|t-1},\Theta)-\frac{\partial h(X_t,\Theta)}{\partial X_t}\at[\Big]{X_t=X_{t|t-1}} X_{t|t-1} , \\
&B_t=\frac{\partial h(X_t,\Theta)}{\partial X_t}\at[\Big]{X_t=X_{t|t-1}}
\end{align}
the affine approximation becomes
\begin{align}
y_t \approx A_t+B_t X_t+\varepsilon_t.
\end{align}
The measurement residuals are computed using the time $t$ futures rates data, $y_t$, and the model implied futures rates. 
\begin{align}
\nu_t = y_t-h(X_t,\Theta).
\end{align}
The residuals have a conditional mean of 0 and conditional variance given by
\begin{align}
S_t:=\mathbb{V}[\nu_t|y_{t-1},...,y_1]=H+B_t P_{t|t-1} B_t'.
\end{align}
In the update step the a priori state estimates are updated using the observed time $t$ data
\begin{align}
&X_{t|t}=X_{t|t-1} +K_t \nu_t, \\
&P_{t|t}=(I-K_t B_t) P_{t|t-1}
\end{align}
where $K_t$ denotes is the optimal Kalman gain matrix given by
\begin{align}
K_t = P_{t|t-1} B_t ' S_t^{-1}.
\end{align}
The resulting Gaussian log-likelihood for a given set of parameters, $\Theta$, is determined by the conditional mean and variance of the innovations $\nu_t$
\begin{align}
l(y_1,...,y_T;\Theta)=- \frac{NT}{2}\log(2\pi) - \frac{1}{2}\sum_{t=1}^T \left(\log |S_t| +\nu_t ' S_t^{-1} \nu_t \right)
\end{align}
To obtain the optimal set of parameters, $\hat{\Theta}$, we maximize the log-likelihood function using the Nelder-Mead algorithm with a function value tolerance of $0.01$. The state variables are required to be stationary under the $P$-measure and we therefore require the eigenvalues of $K^P$ to be positive.
\subsection{Efficiency of the Extended Kalman Filter}
The validity of the results heavily relies on the model parameters as well as underlying state variables being accurately estimated by the extended Kalman filter. Therefore, we construct a simulation study to investigate the efficiency of the extended Kalman filter maximum-likelihood method when estimating the model based on data from futures contracts. We assume that the observed futures follow the AFNS model with parameters equal to the final set of parameters obtained for the full SOFR dataset. The underlying state variables are simulated using the discretized $P$-dynamics
\begin{align*}
X_{t+1}=X_t + K^P (\theta^P - X_t)\Delta t +\sqrt{\Delta t}\Sigma Z_t
\end{align*}
with $Z_t\sim \mathcal{N}(0, I_3)$ and $\Delta t$ set to $1/250$ to reflect the daily observations and the value used in the actual estimation problem. The process is started in the unconditional mean $X_0=\theta^P$.
To reflect the SOFR-estimation the simulated data consists of seven consecutive observations of one-month futures and five consecutive observations of three-month futures. To simplify the study we assume that the contract lengths are 30 and 90 days for all one- and three-month futures. Further, we assume that at all times we observe both the first one- and three-month contract at the beginning of the accrual period, thus we do not have to factor in any rates that have already been accumulated by the contracts. The set of start dates (denoted in ACT/360) for the simulated one- and three-months contracts respectively are therefore $\tau^{1m}=\{0,30/360,60/360,90/360,120/360,150/360,180/360\}$ and $\tau^{3m}=\{0,90/360,180/360,270/360,360/360\}$. The simulated datasets contain 500 observations reflecting approximately two years of daily data. The futures prices are assumed to be observed without error, but rounded to nearest half basis point representing the minimum price fluctuation on CME SOFR futures\footnote{The nearest one and three-month futures contract both have a minimum price fluctuation of $0.25$ basis point. We disregard this minor detail here.}. We base the simulation study on $1.000$ paths. To avoid misspecification the optimization algorithm is started in the true set of parameter values.
\begin{table}[hbt!]
\noindent\makebox[\textwidth]{%
\begin{tabular}{ccccccccc} 
\textbf{Parameter}    & \textbf{True Value} & \textbf{Mean}    & \textbf{SD} & \textbf{5\%}   & \textbf{Q1} & \textbf{Median}  & \textbf{Q3} & \textbf{95\%}   \\ \hline
$k_{11}^P$   & 0.0980     & 0.0763  & 0.1331             & 0.0032  & 0.0247       & 0.0512  & 0.0881       & 0.2072   \\ 
$k_{22}^P$   & 0.5153     & 0.4753  & 0.7928             & 0.0072  & 0.0697       & 0.2855  & 0.5843       & 1.5521   \\ 
$k_{33}^P$   & 2.4486    & 2.1291 & 2.2305             & 0.0327 & 0.3457      & 1.4599 & 3.2279      &   6.5968 \\ 
$\theta_1^P$   & 0.0175     & 0.0093  & 0.0037            & -0.0002  & 0.0031      & 0.0051  & 0.0135       & 0.0277   \\ 
$\theta_2^P$ & -0.0037     & -0.0004 & 0.0018             & -0.0055 & -0.0025      & -0.0006 & 0.0012       & 0.0058   \\ 
$\theta_3^P$ & -0.0012    & -0.0003 & 0.0002             & -0.0024 & -0.0010      & -0.0003 & 0.0003      & 0.0020  \\ 
$\sigma_{11}$   & 0.0054    & 0.0054 & 0.0002             & 0.0051 & 0.0053       & 0.0054 & 0.0055      & 0.0057  \\ 
$\sigma_{22}$   & 0.0062     & 0.0062  & 0.0003             & 0.0059  & 0.0061       & 0.0062  & 0.0063       & 0.0065   \\
$\sigma_{33}$   & 0.0088     & 0.0088  & 0.0003             & 0.0083  & 0.0086       & 0.0088  & 0.0090       & 0.0093   \\ 
$\lambda$    & 2.0284     & 2.0281  & 0.0047             & 2.0206  & 2.0251       & 2.0281  & 2.0310       & 2.0357   \\ \hline
\end{tabular}}
\caption{Estimation results using the Kalman Filter on 500 daily simulated futures observations. The results are based on 1000 simulations.}
\label{ekf1bps}
\end{table}

Focusing on the parameters specific to the physical measure, $K^P$ and $\theta^P$, in table \ref{ekf1bps} we note that these parameters are poorly estimated with fairly large standard errors. As shown in \cite{christensen2013efficient} this is a common issue when estimating Gaussian term structure models and getting a sensible estimate of the physical drift and mean reversion requires a much longer data sample than what is currently available from the SOFR futures market.
However, when turning to the parameters governing the risk-neutral dynamics, $\lambda$ and $\Sigma$, we note that these are estimated close to their respective true values showing no significant bias and a low standard deviation from the true values. This is promising since only the risk-neutral dynamics are relevant when computing the current term structure, while the drift parameters under the physical measure are solely required in e.g. forecasting exercises. In order to produce a valid term structure it is also important that the filtering process is able to accurately reproduce the value of the latent state variable. To test this aspect we compare the true final set of state variables from each simulation with the final filtered set of state variables produced by the Kalman filter. Note that we use the standard Level, Slope and Curve to describe each the three factors of the AFNS model. The results in table \ref{ekfstate} indicate that the filtered state variables closely track the true values with errors of less than one basis point.
\begin{table}[hbt!]
\noindent\makebox[\textwidth]{%
\begin{tabular}{cllllllll}
\multicolumn{1}{l}{\textbf{State Variable}} & \textbf{RMSE} & \textbf{Mean} & \textbf{SD} & \textbf{5\%} & \textbf{Q1} & \textbf{Median} & \textbf{Q3} & \textbf{95\%} \\ \hline
Level                                           & 0.3           & 0.0           & 0.3                         & -0.4         & -0.2                  & 0.0             & 0.2                   & 0.4           \\
Slope                                           & 0.2           & 0.0           & 0.2                         & -0.4         & -0.2                  & 0.0             & 0.2                   & 0.4           \\
Curve                                           & 0.6           & 0.0           & 0.6                         & -1.0         & -0.5                  & 0.0             & 0.4                   & 1.0           \\ \hline
\end{tabular}}
\caption{Statistics on the difference between the true final state variable and the filtered state variable obtained from the Kalman Filter. All values are in basis points.}
\label{ekfstate}
\end{table}

\section{Appendix D: Futures Rate Approximations}
\begin{figure}[hbt!]
\centering
\includegraphics[angle=0, scale=.6]{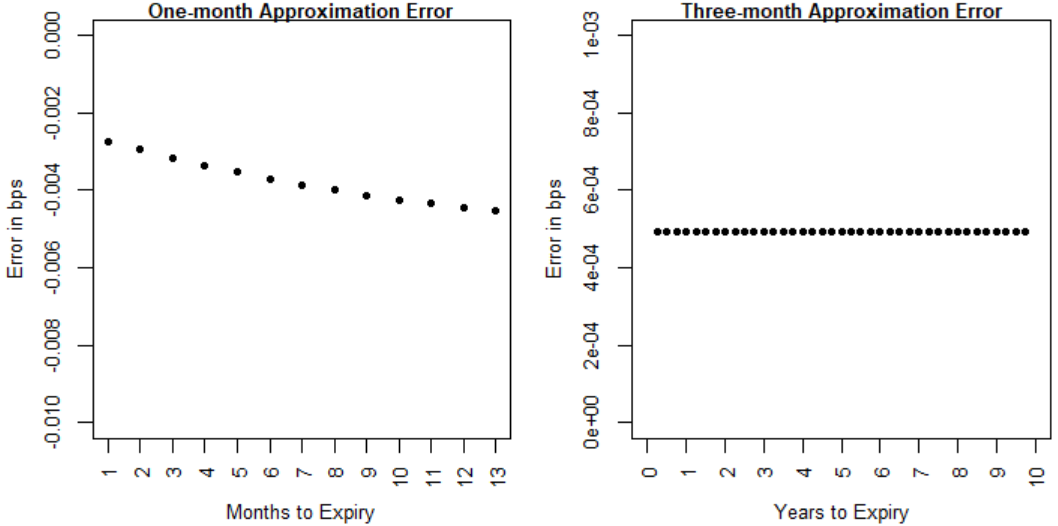}
\caption{Approximation error for one- and three-month futures rates based on the final SOFR parameter estimates.}
\label{fig:fut_err}
\end{figure}

In the following we compare exact futures rates to rates obtained from the approximations presented in section 2 and used in the historical estimation. Since we consider Gaussian term structure models exact expressions for the futures rates of both the one- and three-month contracts exist. The results shown here are based on the three-factor AFNS model but can similarly be obtained for the one- and two-factor Gaussian models.\\
To illustrate the size of the approximation we consider the final set of parameters obtained from the full period of SOFR futures data with the state variable $X_t$ equal to the unconditional mean $\theta^P$. We define the approximation error as the difference between the approximation and the exact futures rate. Figure \ref{fig:fut_err} presents the approximation error covering 13 consecutive one-month and 39 consecutive three-month contracts.
The size of the error is similar across all consecutive contracts and small compared to the minimum price fluctuation of the futures contract. Based on this the approximations do not have any significant impact on the model estimation and resulting term rates.
\subsection{Exact One-Month SOFR Futures Rate}
Recalling that the one month futures rate is given by the discrete average $R(t,T)=\frac{1}{N}\sum_{i=1}^N\frac{1}{d_{i}}\left(\frac{1}{p(t_i,t_{i}+d_{i})}-1 \right)$, we can calculate the mean of the average overnight rate during the reference period as
\begin{align}
f^{1m}(t;S,T)&=\mathbb{E}^Q\left[\frac{1}{N}\sum_{i=1}^N \frac{1}{d_{i}}\left(\frac{1}{p(t_i,t_{i}+d_{i})}-1 \right) |\mathcal{F}_t\right]  \nonumber
\\&=\frac{1}{N}\sum_{i=1}^N \frac{1}{d_{i}}\left(\mathbb{E}^Q\left[\frac{1}{\mathbb{E}^Q\left[e^{-\int_{t_i}^{t_{i}+d_{i}}r_s ds}|\mathcal{F}_{t_i}\right]}|\mathcal{F}_t\right]-1 \right)  \nonumber
\\&=\frac{1}{N}\sum_{i=1}^N \frac{1}{d_{i}}\left(\mathbb{E}^Q\left[\frac{1}{e^{A(t_i,t_{i}+d_{i})+B(t_i,t_{i}+d_{i})'X_{t_i}}}|\mathcal{F}_t\right]-1 \right)   \nonumber
\\&=\frac{1}{N}\sum_{i=1}^N \frac{1}{d_{i}}\left(\mathbb{E}^Q\left[e^{-A(t_i,t_{i}+d_{i})-B(t_i,t_{i}+d_{i})'X_{t_i}} |\mathcal{F}_t\right]-1 \right)  \nonumber
\\&=\frac{1}{N}\sum_{i=1}^N \frac{1}{d_{i}}\left(e^{-A(t_i,t_{i}+d_{i})}e^{-B(t_i,t_{i}+d_{i})'e^{-K^Q (t_i-t)} X_t+\frac{1}{2}B(t_i,t_{i}+d_{i})'\mathbb{V}^{Q}[X_{t_i}|\mathcal{F}_t]B(t_i,t_{i}+d_{i})}-1 \right) .
\end{align}
\subsection{Exact Three-Month SOFR Futures Rate}
To gauge the size of the approximation error of the three-month futures rate using the continuous approximation we follow \cite{henrard2018overnight}. Here it is shown that the three-month futures rate in a Gaussian Heath-Jarrow-Morton setup without approximation is given by 
\begin{align}
f^{3m}(t;S,T)=\frac{1}{T-S}\left(\frac{p(t,S)}{p(t,T)}\prod_{i=1}^n \gamma(t_{i-2},t_{i-1},t_{i-1},t_{n})-1  \right)
\end{align}
with $t_{-1}=t$, $S=t_0 < t_1<,...,<t_{n-1}<t_n=T$ and
\begin{align}
\gamma(s,t,u,v)=\exp\left( \int_s^t \nu (\tau,v)(\nu (\tau,v)-\nu (\tau,u))'d\tau \right)
\end{align}
where $\nu (t,T)$ defines the $p(t,T)$ bond volatility. By application of the multi-dimensional Itô's lemma with $g(t,X_t)=e^{A(t,T)+B(t,T)'X_t}$ we have the following $Q$-dynamics
\begin{align}
dp(t,T)=dg(t,X_t)&=r_t p(t,T)dt + \sum_{i=1}^n \frac{\partial f}{\partial x_i	} \sigma_i dW_t \nonumber \\
&=r_t p(t,T)dt +\sum_{i=1}^n B_i(t,T) e^{A(t,T)+B (t,T)'X_t} \sigma_i dW_t \nonumber\\
&=r_t p(t,T)dt +p(t,T)\sum_{i=1}^n B_i(t,T) \sigma_i dW_t \nonumber\\
&=r_t p(t,T)dt +p(t,T)\nu(t,T) dW_t,
\end{align}
with $\sigma_i$ denoting the row vectors of the volatility matrix, $\Sigma$.
Thus in the three factor AFNS model with a diagonal volatility matrix the $T$-bond volatility is
\begin{align}
\nu(t,T)'=
\begin{pmatrix}
-(T-t)\sigma_{11} \\ -\frac{1-e^{-\lambda(T-t)}}{\lambda}\sigma_{22} \\ \left((T-t)e^{-\lambda(T-t)}-\frac{1-e^{-\lambda(T-t)}}{\lambda}\right) \sigma_{33}
\end{pmatrix}.
\end{align}
The expression for $\gamma(s,t,u,v)$ becomes rather involved for the three-factor model, it is however a scalar that can be analytically computed. 

\end{document}